\documentclass[%
 reprint,
nofootinbib,
 amsmath,amssymb,
 aps,
]{revtex4-2}
\usepackage{graphicx}
\usepackage{dcolumn}
\usepackage{bm}
\usepackage{lineno}
\usepackage{amsmath}
\usepackage{comment}
\usepackage{xcolor}
\usepackage{float}
\usepackage[para, referable]{threeparttablex}
\usepackage[normalem]{ulem}
\usepackage{longtable}

\newcommand{\Fermi}{\emph{Fermi}}
\newcommand{\lat}{{\emph{Fermi}-LAT} }

\newcolumntype{H}{>{\setbox0=\hbox\bgroup}c<{\egroup}@{}}

\newcommand{\measuredsize}{30}

\newcommand{\benchmarksize}{42}
\newcommand{\inclusivesize}{50}
\newcommand{\specialsize}{8}

\newcommand{\allSigEst}{$2-3 \,\sigma$}

\newcommand{\benchmarkSig}{$2.0\,\sigma$}
\newcommand{\benchmarkSigEst}{$\sim 2 \,\sigma$}
\newcommand{\benchmarkMassrange}{$150-230$}

\newcommand{\measplocmaxMass}{8.4}
\newcommand{\consplocmaxMass}{289.4 }
\newcommand{\benchplocmaxMass}{180.5 }
\newcommand{\incplocmaxMass}{$366.5$ }

\newcommand{\measplocmax}{$9.6\times 10^{-2}$}

\newcommand{\benchplocmax}{$2.1\times 10^{-2}$}
\newcommand{\incplocmax}{$9.0\times10^{-4}$}

\newcommand{\meassloc}{$1.3\,\sigma$}

\newcommand{\benchsloc}{$2.0\,\sigma$}
\newcommand{\incsloc}{$3.1\,\sigma$}

\newcommand{\measpglobal}{$2.0\times 10^{-1}$}

\newcommand{\benchpglobal}{$3.1\times 10^{-1}$}
\newcommand{\incpglobal}{$1.1\times 10^{-1}$}

\newcommand{\meassglobal}{$0.9\,\sigma$}

\newcommand{\benchsglobal}{$0.5\,\sigma$}
\newcommand{\incsglobal}{$1.2\,\sigma$}

\newcommand{\benchmarkSigEstTau}{$\sim 2 \,\sigma$}
\newcommand{\benchmarkMassrangeTau}{$30-50$}

\newcommand{\measplocmaxMassTau}{10.6}
\newcommand{\consplocmaxMassTau}{43.8}
\newcommand{\benchplocmaxMassTau}{43.8}
\newcommand{\incplocmaxMassTau}{$43.8$ }

\newcommand{\measplocmaxTau}{$9.6\times 10^{-2}$}

\newcommand{\benchplocmaxTau}{$2.3\times 10^{-2}$}
\newcommand{\incplocmaxTau}{$1.1\times10^{-3}$}

\newcommand{\measslocTau}{$1.3\,\sigma$}

\newcommand{\benchslocTau}{$2.0\,\sigma$}
\newcommand{\incslocTau}{$3.1\,\sigma$}

\newcommand{\measpglobalTau}{$3.2\times 10^{-1}$}

\newcommand{\benchpglobalTau}{$4.2\times 10^{-1}$}
\newcommand{\incpglobalTau}{$1.3\times 10^{-1}$}

\newcommand{\meassglobalTau}{$0.5\,\sigma$}

\newcommand{\benchsglobalTau}{$0.2\,\sigma$}
\newcommand{\incsglobalTau}{$1.1\,\sigma$}

\makeatletter
\renewenvironment{widetext@grid}{%
  \par\ignorespaces
  \setbox\widetext@top\vbox{%
   \vskip15\p@
   \hb@xt@\hsize{%
    \leaders\hrule\hfil
    \vrule\@height6\p@
   }%
   \vskip6\p@
  }%
  \setbox\widetext@bot\hb@xt@\hsize{%
    \vrule\@depth6\p@
    \leaders\hrule\hfil
  }%
  \onecolumngrid
  \let\set@footnotewidth\set@footnotewidth@ii
}{%
  \par
  \twocolumngrid\global\@ignoretrue
  \@endpetrue
}%

\makeatother

\begin{document}

\def\reff@jnl#1{{\rm#1}} 

\def\araa{\reff@jnl{ARA\&A}}             
\def\aj{\reff@jnl{AJ}}                   
\def\apj{\reff@jnl{ApJ}}                 
\def\apjl{\reff@jnl{ApJ}}                
\def\apjs{\reff@jnl{ApJS}}               
\def\apss{\reff@jnl{Ap\&SS}}             
\def\aap{\reff@jnl{A\&A}}                
\def\aapr{\reff@jnl{A\&A~Rev.}}          
\def\aaps{\reff@jnl{A\&AS}}              
\def\baas{\ref@jnl{BAAS}}               
\def\jcap{\reff@jnl{J. Cosmology Astropart. Phys.}}
\def\jrasc{\reff@jnl{JRASC}}             
\def\memras{\reff@jnl{MmRAS}}            
\def\mnras{\reff@jnl{MNRAS}}             
\def\na{\reff@jnl{New A}}                
\def\nar{\reff@jnl{New A Rev.}}          
\def\pra{\reff@jnl{Phys.~Rev.~A}}        
\def\prb{\reff@jnl{Phys.~Rev.~B}}        
\def\prc{\reff@jnl{Phys.~Rev.~C}}        
\def\prd{\reff@jnl{Phys.~Rev.~D}}        
\def\pre{\reff@jnl{Phys.~Rev.~E}}        
\def\prl{\reff@jnl{Phys.~Rev.~Lett.}}    
\def\pasa{\reff@jnl{PASA}}               
\def\pasp{\reff@jnl{PASP}}               
\def\pasj{\reff@jnl{PASJ}}               
\def\qjras{\reff@jnl{QJRAS}}             
\def\physrep{\reff@jnl{Phys.~Rep.}}   
\let\astap=\aap
\let\apjlett=\apjl
\let\apjsupp=\apjs
\let\applopt=\ao

\preprint{APS/123-QED}

\title{Legacy Analysis of Dark Matter Annihilation from the Milky Way Dwarf Spheroidal Galaxies with 14 Years of \Fermi{}-LAT Data}

\author{Alex McDaniel}
\email{armcdan@clemson.edu}
\affiliation{Department of Physics and Astronomy, 
Clemson University,
Clemson, SC, 29631, USA}

\author{Marco Ajello}
\affiliation{Department of Physics and Astronomy, 
Clemson University,
Clemson, SC, 29631, USA}

\author{Christopher M. Karwin}
\affiliation{NASA Postdoctoral Program Fellow, NASA Goddard Space Flight Center, 
Greenbelt, MD, 20771, USA}

\author{Mattia Di Mauro}
\affiliation{Dipartimento di Fisica, Universit\'a di Torino, Via P. Giuria 1, 10125 Torino, Italy}
\affiliation{Istituto Nazionale di Fisica Nucleare, Sezione di Torino, Via P. Giuria 1, 10125 Torino, Italy
}%
\author{Alex Drlica-Wagner}
\affiliation{
Fermi National Accelerator Laboratory, P.O. Box 500, Batavia, IL 60510, USA
}%
\affiliation{Kavli Institute for Cosmological Physics, University of Chicago, Chicago, IL 60637, USA}
\affiliation{Department of Astronomy and Astrophysics, University of Chicago, Chicago IL 60637, USA}

\author{Miguel S\`{a}nchez-Conde}
\affiliation{
Instituto de F\'isica Te\'orica, IFT UAM-CSIC, Departamento de F\'isica Te\'orica,
Universidad Aut\'onoma de Madrid, ES-28049 Madrid, Spain
}%

\date{\today}

\begin{abstract}
The Milky Way dwarf spheroidal satellite galaxies (dSphs) are particularly intriguing targets to search for gamma rays from Weakly Interacting Massive Particle (WIMP) dark matter (DM) annihilation or decay. They are nearby, DM-dominated, and lack significant emission from standard astrophysical processes. 
Previous studies using the \Fermi{}-Large Area Telescope (LAT) of DM-induced emission from dSphs have provided the most robust and stringent constraints on the DM annihilation cross section and mass. We report here an analysis of the Milky Way dSphs using over 14 years of LAT data along with an updated census of dSphs and $J$-factor estimates. While no individual dSphs are significantly detected, we do find slight excesses with respect to background at the $\gtrsim 2 \,\sigma$ local significance level in both tested annihilation channels ($b\bar{b}$, $\tau^+\tau^-$) for seven of the dSphs. We do not find a significant DM 
signal from the combined likelihood analysis of the dSphs ($s_{global}\sim 0.5 \sigma$), yet a marginal local excess relative to background at a \allSigEst{} level is observed at a DM mass of $M_{\chi}=$ \benchmarkMassrange{} GeV ($M_{\chi}=$ \benchmarkMassrangeTau{} GeV) for DM annihilation into $b \bar{b}$ ($\tau^+\tau^-$). 
Given the lack of a significant detection, we place updated constraints on the $b\bar{b}$ and $\tau^+\tau^-$ annihilation channels that are generally consistent with previous recent results. As in past studies, tension is found with some WIMP DM interpretations of the Galactic Center Excess (GCE), though the limits are consistent with other interpretations given the uncertainties of the Galactic DM density profile and GCE systematics. Based on conservative assumptions of improved sensitivity with increased {\it Fermi}-LAT exposure time and moderate increases in the sample of Milky Way dSphs, we project that the local \benchmarkSigEst{} signal, if real, could approach the $\sim 4 \,\sigma$ local confidence level with additional $\sim 10$ years of observation. 
\end{abstract}

\maketitle


\section{Introduction}

Evidence from a wide body of astrophysical and cosmological observations suggests that dark matter (DM) contributes $\sim 84\%$ of the matter budget of the Universe \cite{planck2018}; however, the fundamental nature of DM has yet to be determined. Several experimental searches including collider searches, direct detection, and indirect detection have been employed, though no definitive signatures of DM interactions have been found \cite{bertone, bergstrom,arcadi2018}. Nonetheless, the varied DM searches provide constraints on DM properties, narrow the allowed parameter space for proposed models, and, in some cases, provide putative DM induced signals. 

Indirect detection techniques are a powerful tool to explore DM properties. They search for DM signatures in the astrophysical systems where the presence of DM can be independently measured and quantified from the dynamics of the system, and they have provided some of the most stringent constraints on well-motivated DM models. In particular, models of annihilating DM, such as Weakly Interacting Massive Particles (WIMPs), predict an array of multiwavelength and multimessenger signatures \cite{jungman, bergstrom2000,bertone, colafrancesco}. Gamma-ray observations stand out as one of the most powerful of these signatures to probe DM, as evidenced by a number of previous works presenting DM constraints (e.g., \cite{abdo2010, ackermann2011,ackermann2014, ackermann2015, drlicawagner2015_dm, geringerSameth2015,albert2017,calore2018, hoof2020, diMauro2021,ackermann2015_virgo, thorpeMorgan2021, dimauro2023_cluster,li2016, albert_m31, karwin2021, burns2021,arcadi2018, dimauro2022}). The expected gamma-ray flux from DM annihilation is given by \cite{bergstrom1998}:

\begin{equation}\label{eq:DMflux}
\frac{d\Phi_{\chi}}{dE} = J \times \frac{1}{4 \pi} \frac{\left<\sigma v\right>}{2M_{\chi}^2}\sum_i \beta_i\frac{dN_i}{dE},
\end{equation}
where $M_{\chi}$ is the rest mass of the DM particle, $\left<\sigma v\right>$ is the DM annihilation cross section, $\frac{dN}{dE}$ is the gamma-ray spectrum per annihilation for a given annihilation channel, and the sum is performed over each of the annihilation channels with branching ratios $\beta_i$. For annihilation through a single channel, $\beta_i = 1$. The quantity $J$ is the so-called ``$J$-factor'', which is the integral of the DM density squared ($\rho_{\chi}^2$) along the line of sight ($\ell$, l.o.s.) and the solid angle ($\Delta \Omega$):
\begin{equation}
J = \int_{\Delta \Omega}\int_{l.o.s.}\rho_{\chi}^2d\ell d\Omega.
\end{equation}

Gamma-ray searches for DM encompass a wide range of targets, including the extragalactic gamma-ray background \cite{abazajian2010_egb,abdo2010_egb,bringmann2014,ajello2015_egb,fermi2014_egb,dimauro2015_egb,ando2013}, galaxy groups \cite{lisanti2018a, lisanti2018} and clusters \cite{ackermann2015_virgo, thorpeMorgan2021, dimauro2023_cluster}, the nearby Andromeda galaxy \cite{li2016, albert_m31,karwin2019, karwin2021, burns2021}, dark subhalos \cite{fermi_subhalos2012, berlin2014, schoonenberg2016, coronadoBlazquez2019a, coronadoBlazquez2019b, coronadoBlazquez2020, coronadoBlazquez2022}, dwarf galaxies \cite{abdo2010, ackermann2011, ackermann2015, drlicawagner2015_dm, geringerSameth2015,albert2017,calore2018, hoof2020, diMauro2021}, and the Milky Way Galactic Center (GC) \cite{goodenoughHooper2009,HooperGoodenough2011,GordonMacias2013,calore2015,abazajian2016,daylan2016,ajello2016_GCE,karwin2017_GCE,dimauro2021_GCE, FermiReview}. Although no conclusive DM signal has been found, dwarf galaxies -- specifically Milky Way dwarf spheroidal (dSph) satellite galaxies -- have yielded some of the most stringent and robust DM constraints derived from gamma-ray observations.  dSphs stand out as particularly powerful targets for DM searches due to their proximity, low astrophysical backgrounds, and high DM concentrations. For $M_{\chi}\lesssim 100$ GeV, constraints from dSphs reach below the canonical thermal relic cross-section ($\sim 2\times 10^{-26}$ cm$^{3}$ s$^{-1}$ \cite{steigman}), and extend down to $10^{-27}-10^{-26}$ cm$^{3}$ s$^{-1}$ depending on the mass and annihilation channel under consideration.

Analyses of dSphs are furthermore of significance as they offer a critical test of the purported DM origin of the gamma-ray Galactic center excess (GCE)
\cite{goodenoughHooper2009,HooperGoodenough2011,GordonMacias2013,calore2015,abazajian2016,daylan2016,ajello2016_GCE,karwin2017_GCE,dimauro2021_GCE, cholis2022}. These studies have typically found that the GCE can be well-modeled by annihilating DM with a mass of $M_{\chi}\sim 40$ GeV and annihilation cross section of $\left<\sigma v\right>= 1-3 \times 10^{-26}$  cm$^{3}$ s$^{-1}$ for annihilation through the $b\bar{b}$ channel, or $M_{\chi}\sim 10$ GeV and $\left<\sigma v\right>= 2\times 10^{-27} - 2 \times 10^{-26}$  cm$^{3}$ s$^{-1}$ for annihilation through the $\tau^+\tau^-$ channel. As further pressure is placed on the WIMP paradigm, improved constraints continue to be of great value in confining the allowed regions of the DM parameter space. 

Typically, studies of DM induced gamma-ray emission from dSphs include a combined likelihood analysis over a sample of targets in order to improve sensitivity. Early studies of DM induced gamma-ray emission used a small sample of classical dSphs (e.g., 7 in \cite{geringersameth2011} and 10 in  \cite{ackermann2011}). Over the years, the collection of known dSphs has grown significantly from the early population of dSphs  \cite{mcconnachie2012} and now comprises over 50 systems \cite{drlica_wagner2020}. The growth of the known dSph population has resulted from optical sky surveys including the Sloan Digital Sky Survey (SDSS) \cite{willman2005_a, willman2005_b, belokurov2006,belokurov2007,belokurov2008,belokurov2009,belokurov2010, grillmair2009, kim2015_peg, zucker2006_ursa, zucker2006_canes, sakamoto2006, irwin2007, walsh2007}, Dark Energy Survey (DES) \cite{ bechtol2015,drlicawagner2015_desy2, koposov2015, kimJerjen2015_horo}, other Dark Energy Camera (DECam) surveys \cite{martin2015, drlicawagner2016, koposov2018, torrealba2018, mau2020, cerny2021,cerny2022,cerny2023,cerny2023b}, Panoramic Survey Telescope and Rapid Response System (Pan-STARRS) \cite{laevens2015_a,  laevens2015_b}, ATLAS \cite{ torrealba2016_crater, torrealba2016_Aq2}, the Hyper Suprime-Cam Subaru Strategic Program \cite{homma2016, homma2018, homma2019}, and {\it Gaia} \cite{torrealba2019}. The population of known dSphs is continually growing as additional optical data are obtained and analyzed (e.g., \citep{drlicawagner2021, drlicawagner2022}), and future observations with the Vera C.\ Rubin Observatory are expected to accelerate the discovery of new systems \cite{tollerud2008,hargis2014,jethwa2018, newton2018, nadler2020, manwadkar2022}.

Beyond the detection of new dSphs, reliable estimates of their DM content are essential for studies of DM annihilation. Improved stellar kinematic data along with the increasing population of known dSphs has allowed for better measurements of dSphs $J$-factors and the derivation of more reliable scaling relationships between the $J$-factors and dSphs kinematic and photometric properties \cite{drlicawagner2015_dm,evans2016,paceStrigari2019}.

The \lat has now observed the gamma-ray sky for over 15 years and has released updated source catalogs (e.g., 4FGL-DR3 \cite{dr3})\footnote{Catalogs are publicly available on the HEASARC website, available at \url{https://heasarc.gsfc.nasa.gov/docs/heasarc/biblio/pubs/fermi_ycat.html}.}. This improved data set is expected to increase the sensitivity to faint gamma-ray emission in Milky Way dSphs. This paper presents the most recent dSph DM search with the {\it Fermi}-LAT, incorporating a larger dSph sample, a longer {\it Fermi}-LAT exposure, the inclusion of one of the more recent \Fermi{}-LAT source catalogs (4FGL-DR3), and updated $J$-factor measurements and scaling relations. 

This paper is organized as follows. In Section \ref{sec:sample} we present the sample of dSphs and $J$-factors used in this analysis. In Section \ref{sec:fermi} we discuss the \lat data used and the analysis procedure. We present the results of the analysis in Section \ref{sec:results} and discuss the capability of improved detection given additional \Fermi-LAT data. Finally, we conclude in Section \ref{sec:conclusion}.

\section{Sample Selection}\label{sec:sample}
The known population of dSphs has expanded rapidly over the last decade. Since the detection, confirmation, and measurement of the dSphs require a variety of observational resources, dSphs can be roughly grouped into at least four categories based on their observational statuses.  These are i) dSphs with confirmed DM content and measured $J-$factors, ii) dSphs with confirmed DM content but without measured $J$-factors, iii) confirmed dSphs without confirmed DM content  iv) stellar systems that are candidates to be dSphs. The first three categories  together constitute the class of ``confirmed’’ dSphs, while the last category is the set of ``probable’’ or candidate dSphs. For the confirmed dSphs with confirmed DM content, the DM content has been inferred directly from stellar kinematic measurements. In the case of the probable dSphs, the systems do not have measured stellar kinematics, but have photometric properties consistent with previously known dSphs. The probable dSphs represent important new targets for indirect DM searches with gamma rays.

The sample of dSphs used in this analysis is taken from the study of Milky Way satellites presented in \cite{drlica_wagner2020}, which performs a census of dSphs covering $\sim75\%$ of the high-Galactic-latitude sky. Included in \cite{drlica_wagner2020} are 57 confirmed or probable dSphs. We exclude several systems from our analysis due to the presence of astrophysical gamma-ray emission and/or dynamical measurements as we describe in detail here. We exclude the Large and Small Magellanic Clouds due to their extended and complex nature and the presence of astrophysical gamma-ray emission which is difficult to disentangle from potential DM contributions (for dedicated studies of these systems, see \cite{buckley2015_LMC, caputo2016_SMC}). We include several dSphs that lack measured $J$-factors, in which case we use predicted $J$-factors based on the scaling relations reported in \cite{paceStrigari2019} (see Section \ref{sec:Jfactors}). There are a few systems where spectroscopic measurements exist, but indicate particularly low velocity dispersions and only upper limits on the dispersions (and thus the $J$-factors) are obtained. Moreover, the estimated $J$-factors based on the scaling relations for these systems are inconsistent with the upper limits. Since the observations provide only upper limits rather than a resolved $J$-factor measurement needed for the analysis, and the estimated $J$-factors are in conflict with the upper limits, these systems are not included in our analysis. These systems include Hydra II, Segue 2, Triangulum II, and Tucana III \cite{paceStrigari2019, kirby2013}. We also exclude sources that lie within the 95\% confidence radius of a 4FGL source with an astrophysical counterpart. This criterion affects only Sculptor, which has coordinates that lie within the 95\% confidence radius of a 4FGL source (4FGL J0059.5-3338) classified based on multiwavelength associations as a flat-spectrum radio quasar (FSRQ) in the 4FGL-DR3 catalog. 

Of the remaining \inclusivesize{} targets, we flag \specialsize{} as being ``Special'' cases indicating that they should be considered with caution as a part of the analysis. Some dSphs in the sample show evidence of non-equilibrium dynamics and tidal-stripping and are thus flagged as ``Special'' cases. These are Willman 1 \cite{willman2011} and Bo\"{o}tes III \cite{carlin2009, carlin2018}\footnote{Tucana III would also fall into this criteria  \cite{drlicawagner2015_desy2, li2018}, though it has already been excluded due to the unresolved velocity dispersion.}. Sagittarius is flagged due in part to its relatively low Galactic latitude (cf. \cite{albert2017}), likely non-equilibrium dynamical state \cite{evans2022}, as well as its proximity to a nearby gamma-ray source. Indeed, the coordinates for Sagittarius lie just outside the 95\% radius of the globular cluster M54 (4FGL J1855.1-3025), which was first included in 4FGL-DR3. Recent dedicated studies have presented possible scenarios for both DM and millisecond pulsar origins of the gamma-ray emission in the vicinity of Sagittarius and the M54 globular cluster, though neither explanation has been confirmed \cite{crocker2022,evans2022}. We cross-matched the sample with multiwavelength catalogs and flag several dSphs that lie in the vicinity ($<0.1^{\circ}$) of blazars or blazar candidate sources as listed in the BZCat \cite{bzcat}, CRATES \cite{crates}, or WIBRaLS \cite{dabrusco2014} catalogs. However, there is not a 4FGL catalog association to these sources so they are not excluded from the analysis (as in the case of Sculptor). The dSphs that are flagged by this criteria are Bo\"{o}tes I, Crater II, Antlia II, Horologium II, and Virgo I.

We divide the sample of dSphs into subsets for detailed consideration. These subsets are:
\begin{itemize}
\item {\bf Measured} (\measuredsize{}): Only dSphs with measured $J$-factors, excluding ``Special'' cases.
\item {\bf Benchmark} (\benchmarksize{}): dSphs with measured or estimated $J$-factors, excluding ``Special'' cases.
\item {\bf Inclusive} (\inclusivesize{}): dSphs with measured or estimated $J$-factors, including ``Special'' cases.
\end{itemize}

These subsamples are also shown in Figure \ref{fig:Jfactors}. The classification of targets as confirmed or probable dSphs involves some level of subjectivity. In some instances, the velocity dispersion of the system may be marginally resolved and a $J$-factor measurement can be obtained despite the inability to conclusively classify the systems as dSphs (e.g., Grus I \cite{walker2016, chiti2022}, Draco II \cite{longeard2018}, see also \cite{paceStrigari2019}). On the other hand, systems may have kinematic data and properties indicating that they are indeed dSphs; however, the spectroscopic data are not sufficient to determine a reliable measured $J$-factor, or the $J$-factor simply has not been determined from the data yet. In this analysis, we generally adopt the confirmed versus probable classification from \cite{drlica_wagner2020}, except for Sagittarius II which we designate as a `probable' dSph rather than `confirmed' dSph due to recent studies indicating that it has a relatively low mass-to-light ratio \cite{longeard2020}. The sample of dSphs is listed in Table \ref{tab:sample}, along with relevant properties. These include several parameters taken from \cite{drlica_wagner2020} such as the RA and Dec coordinates used in the analysis, the derived heliocentric distance, the azimuthally averaged physical half-light radius ($r_{1/2}$), and the published absolute V-band magnitude ($M_V$). Also included are the $J$-factor and uncertainty for each dSph, along with an indicator of whether it is a measured $J$-factor or a scaling relation estimate and an indicator of whether it is considered a confirmed or probable dSph.

\newpage
{\setlength{\LTcapwidth}{\textwidth}
\def\arraystretch{1.2}
\setlength{\tabcolsep}{1pt}
\begin{longtable*}[e]{@{\extracolsep{\fill}}lccccccccc}
\hline\hline
(1)&(2)&(3)&(4)&(5)&(6)&(7)&(8)&(9)&(10)\\
Name & R.A. (J2000) & Decl. (J2000)  & Distance & $r_{1/2}$ & $M_V$ & $\sigma_{\mathrm{l.o.s.}}$ & log$_{10}J\pm\sigma_J$ & Method & Status \\
 & [deg]& [deg] & [kpc] & [pc] & [mag] & [km s$^{-1}$] & [log$_{10}\mathrm{GeV^2 \ cm^{-5}}$]& [M/K/P] & [C/P]\\
\hline
\multicolumn{10}{c}{dSphs with Measured $J$-factors}\\
\hline
Aquarius II & 338.48 & -9.33 & 108.0 & 125 & -4.4 & 4.7$^{a}$ & $17.80 \pm 0.55^{a}$ & M &C \\
Bo\"{o}tes II & 209.51 & 12.86 & 42.0 & 39 & -2.94 & 2.9$^{a}$ & $18.30 \pm 0.95^{a}$ & M &C \\
Canes Venatici I & 202.01 & 33.55 & 218.0 & 338 & -8.8 & 7.6 & $17.42 \pm 0.16$ & M &C \\
Canes Venatici II & 194.29 & 34.32 & 160.0 & 55 & -5.17 & 4.7 & $17.82 \pm 0.47$ & M &C \\
Carina & 100.41 & -50.96 & 105.0 & 248 & -9.43 & 6.4 & $17.83 \pm 0.10$ & M &C \\
Carina II & 114.11 & -58.0 & 36.0 & 77 & -4.5 & 3.4 & $18.25 \pm 0.55$ & M &C \\
Coma Berenices & 186.75 & 23.91 & 44.0 & 57 & -4.38 & 4.7 & $19.00 \pm 0.35$ & M &C \\
Draco & 260.07 & 57.92 & 76.0 & 180 & -8.71 & 9.1 & $18.83 \pm 0.12$ & M &C \\
Draco II & 238.17 & 64.58 & 22.0 & 17 & -0.8 & 3.4 & $18.93 \pm 1.54$ & M &P \\
Eridanus II & 56.09 & -43.53 & 380.0 & 158 & -7.21 & 7.1 & $16.60 \pm 0.90$ & M &C \\
Fornax & 39.96 & -34.5 & 147.0 & 707 & -13.46 & 10.6 & $18.09 \pm 0.10$ & M &C \\
Grus I & 344.18 & -50.18 & 120.0 & 21 & -3.47 & 4.5 & $16.50 \pm 0.80$ & M &P \\
Hercules & 247.77 & 12.79 & 132.0 & 120 & -5.83 & 3.9 & $17.37 \pm 0.53$ & M &C \\
Horologium I & 43.88 & -54.12 & 79.0 & 31 & -3.55 & 5.9 & $19.00 \pm 0.81$ & M &C \\
Hydrus I & 37.39 & -79.31 & 28.0 & 53 & -4.71 & 2.7$^b$ & $18.33 \pm 0.36^b$ & M &C \\
Leo I & 152.11 & 12.31 & 254.0 & 226 & -11.78 & 9.0 & $17.64 \pm 0.13$ & M &C \\
Leo II & 168.36 & 22.15 & 233.0 & 165 & -9.74 & 7.4 & $17.76 \pm 0.20$ & M &C \\
Leo IV & 173.24 & -0.55 & 154.0 & 104 & -4.99 & 3.4 & $16.40 \pm 1.08$ & M &C \\
Leo V & 172.79 & 2.22 & 178.0 & 39 & -4.4 & 4.9 & $17.65 \pm 0.97$ & M &C \\
Pegasus III & 336.1 & 5.41 & 215.0 & 42 & -3.4 & 7.9 & $18.30 \pm 0.93$ & M &C \\
Pisces II & 344.63 & 5.95 & 182.0 & 48 & -4.22 & 4.8 & $17.30 \pm 1.04$ & M &C \\
Reticulum II & 53.92 & -54.05 & 30.0 & 31 & -3.88 & 3.4 & $18.90 \pm 0.38$ & M &C \\
Sagittarius II & 298.16 & -22.07 & 69.0 & 32 & -5.2 & 2.7$^c$ & $17.35 \pm 1.36^d$ & M &P \\
Segue 1 & 151.75 & 16.08 & 23.0 & 20 & -1.3 & 3.1 & $19.12 \pm 0.53$ & M &C \\
Sextans & 153.26 & -1.61 & 86.0 & 345 & -8.72 & 7.1 & $17.73 \pm 0.12$ & M &C \\
Tucana II & 342.98 & -58.57 & 58.0 & 165 & -3.8 & 7.3 & $18.97 \pm 0.54$ & M &C \\
Tucana IV & 0.73 & -60.85 & 48.0 & 128 & -3.5 & 4.3$^e$ & $18.40 \pm 0.55^e$ & M &C \\
Ursa Major I & 158.77 & 51.95 & 97.0 & 151 & -5.12 & 7.3 & $18.26 \pm 0.28$ & M &C \\
Ursa Major II & 132.87 & 63.13 & 32.0 & 85 & -4.25 & 7.2 & $19.44 \pm 0.40$ & M &C \\
Ursa Minor & 227.24 & 67.22 & 76.0 & 272 & -9.03 & 9.3 & $18.75 \pm 0.12$ & M &C \\
\hline
\multicolumn{10}{c}{dSphs with Estimated $J$-factors}\\
\hline
Bo\"{o}tes IV & 233.69 & 43.73 & 209.0 & 277 & -4.53 & -- & $17.25 \pm 0.60$ & P & P \\
Carina III & 114.63 & -57.9 & 28.0 & 30 & -2.4 & 5.6$^f$ & $19.70 \pm 0.60$ & K & C \\
Centaurus I & 189.59 & -40.9 & 116.0 & 76 & -5.55 & -- & $18.14 \pm 0.60$ & P & P \\
Cetus II & 19.47 & -17.42 & 30.0 & 17 & 0.0 & -- & $19.10 \pm 0.60$ & P & P \\
Cetus III & 31.33 & -4.27 & 251.0 & 44 & -2.5 & -- & $17.30 \pm 0.60$ & P & P \\
Columba I & 82.86 & -28.01 & 183.0 & 98 & -4.2 & -- & $17.60 \pm 0.60$ & P & P \\
Grus II & 331.02 & -46.44 & 53.0 & 92 & -3.9 & -- & $18.40 \pm 0.60$ & P & P \\
Phoenix II & 355.0 & -54.41 & 83.0 & 21 & -3.3 & -- & $18.30 \pm 0.60$ & P & C \\
Pictor I & 70.95 & -50.29 & 114.0 & 18 & -3.45 & -- & $18.00 \pm 0.60$ & P & P \\
Pictor II & 101.18 & -59.9 & 46.0 & 47 & -3.2 & -- & $18.83 \pm 0.60$ & P & P \\
Reticulum III & 56.36 & -60.45 & 92.0 & 64 & -3.3 & -- & $18.20 \pm 0.60$ & P & P \\
Tucana V & 354.35 & -63.27 & 55.0 & 16 & -1.6 & -- & $18.90 \pm 0.60$ & P & P \\
\hline
\multicolumn{10}{c}{Special Cases} \\
\hline
Antlia II & 143.89 & -36.77 & 132.0 & 2301 & -9.03 & 5.7$^g$ & $16.50 \pm 0.60$ & K & C \\
Bo\"{o}tes I & 210.02 & 14.51 & 66.0 & 160 & -6.02 & 4.9 & $18.17 \pm 0.30$ & M & C \\
Bo\"{o}tes III & 209.3 & 26.8 & 47.0 & 289 & -5.75 & -- & $18.65 \pm 0.60$ & P & C \\
Crater II & 177.31 & -18.41 & 117.5 & 1066 & -8.2 & 2.7$^h$ & $15.35 \pm 0.26^d$ & M & C \\
Horologium II & 49.11 & -50.05 & 78.0 & 33 & -2.6 & -- & $18.40 \pm 0.60$ & P & P \\
Sagittarius & 283.83 & -30.55 & 26.7 & 1565 & -13.5 & -- & $19.60 \pm 0.20^i$ & M & C \\
Virgo I & 180.04 & -0.68 & 91.0 & 30 & -0.33 & -- & $18.10 \pm 0.60$ & P & P \\
Willman 1 & 162.34 & 51.05 & 38.0 & 20 & -2.53 & 4.5 & $19.53 \pm 0.50$ & M & C \\
\hline\hline
\caption{properties of the 50 dSphs used in this analysis. The first section lists the 30 dSphs that have measured $J$-factors and constitute the Measured sample. The second section lists an additional 12 dSphs that have only $J$-factor estimates and together with the Measured section constitute the Benchmark sample. The third section lists the Special cases. The Inclusive sample comprises all dSphs listed in the table. Column descriptions: (1) source name (2) right ascension (3) declination (4) heliocentric distance (5) half-light radius (6) absolute V-band magnitude (7) $J$-factor (8) log-uncertainty on the $J$-factor (9) method used to determine the $J$-factor (10) classification status as a confirmed (C) vs probable (P) dSph. The $J$-factor methods are either to adopt the measured $J$-factors from \cite{paceStrigari2019} whenever available (M = Measured) or estimate them using the kinematic scaling relation (K = Kinematic, Eq. \ref{eq:kinematic}) or photometric scaling relation (P = Photometric, Eq. \ref{eq:photmetric}) derived in \cite{paceStrigari2019}. For $J$-factors predicted from scaling relations we assume an error of 0.6 dex. Columns references: (2-6) \cite{drlica_wagner2020}, (8,9) \cite{paceStrigari2019} unless indicated by a footnote: (a)  \cite{bruce2023}; (b)  \cite{koposov2018}; (c) \cite{longeard2020}; (d) \cite{boddy2020}; (e) \cite{simon2020}; (f)  \cite{li2018_car3}; (g)  \cite{torrealba2019}; (h) \cite{caldwell2017}; (i) \cite{evans2022}.}
\label{tab:sample}
\end{longtable*}

\subsection{Astrophysical $J$-Factors}\label{sec:Jfactors}
The gamma-ray flux from DM annihilation is directly proportional to the astrophysical $J$-factor (see Equation \ref{eq:DMflux}). Thus, the derived constraints rely on reliable measurements or estimations of this value and characterization of its uncertainties. In this analysis, we primarily use the measured $J$-factor values and uncertainties with an integration angle of $0.5^{\circ}$ reported in \cite{paceStrigari2019}. For some dSphs, the $J$-factors were not included in \cite{paceStrigari2019} and updated measurements are available in the literature, in which case those $J$-factor measurements are used (see Table \ref{tab:sample} for detailed references). For dSphs where no measured $J$-factor is available we employ the spectroscopic and photometric scaling relations presented in \cite{paceStrigari2019}. These take the form:

\begin{widetext}
\begin{eqnarray}
 \text{Kinematic:}\:   \frac{J(0.5^{\circ})}{\mathrm{GeV}^2 \mathrm{cm}^{-5}} && = 10^{17.87}\left(\frac{\sigma_{\mathrm{l.o.s}}}{5\,\mathrm{km s}^{-1}}\right)^{4}
   \left(\frac{d}{100\,\mathrm{kpc}}\right)^{-2}\left(\frac{r_{1/2}}{100\,\mathrm{pc}}\right)^{-1}
\label{eq:kinematic} \end{eqnarray}
\begin{eqnarray}
  \text{Photometric:}\:  \frac{J(0.5^{\circ})}{\mathrm{GeV}^2 \mathrm{cm}^{-5}} && = 10^{18.17}\left(\frac{L_V}{10^4L_{\odot}}\right)^{0.23}
   \left(\frac{d}{100\,\mathrm{kpc}}\right)^{-2}\left(\frac{r_{1/2}}{100\,\mathrm{pc}}\right)^{-0.5}
\label{eq:photmetric}\end{eqnarray}
\end{widetext}

 \noindent where $d$ is the distance in kpc, $\sigma_{\mathrm{l.o.s.}}$ is the line of sight velocity dispersion in km s$^{-1}$, $r_{1/2}$ is the half-light radius in pc, and $L_V$ is the V-band luminosity in units of solar luminosity ($L_{\odot}$).
Values for the distances and relevant spectroscopic and photometric parameters that define the scaling relations are taken from \cite{drlica_wagner2020}. If there is not a measured $J$-factor, but spectroscopic data are available, the kinematic scaling relation (Equation \ref{eq:kinematic}) is preferred over the photometric scaling relation (Equation \ref{eq:photmetric}). For dSphs without available kinematic data, we use the photometric relation to estimate the $J$-factor. In all instances where estimated $J$-factors are used we adopt an uncertainty of 0.6 dex. This is an assumed value intended to represent the expected measurement uncertainty on the $J$-factors after kinematic observations. The value  of 0.6 dex is only slightly higher than the mean uncertainty of the measured dSphs ($\sim 0.5$ dex), and is the nominal value used in \cite{albert2017}, allowing for closer comparison with the results of that work.
The $J$-factors used in this analysis for each dSph are shown in Figure \ref{fig:Jfactors}, and are also listed in Table \ref{tab:sample}. Previous works such as \cite{albert2017} also used a scaling relation to derive $J$-factors for targets lacking measured stellar kinematics. However, those relations were based solely on distance, whereas the improved scaling relations used here offer a more reliable estimate of the $J$-factor \cite{paceStrigari2019}.

        \begin{figure*}
            \includegraphics[width=\linewidth]{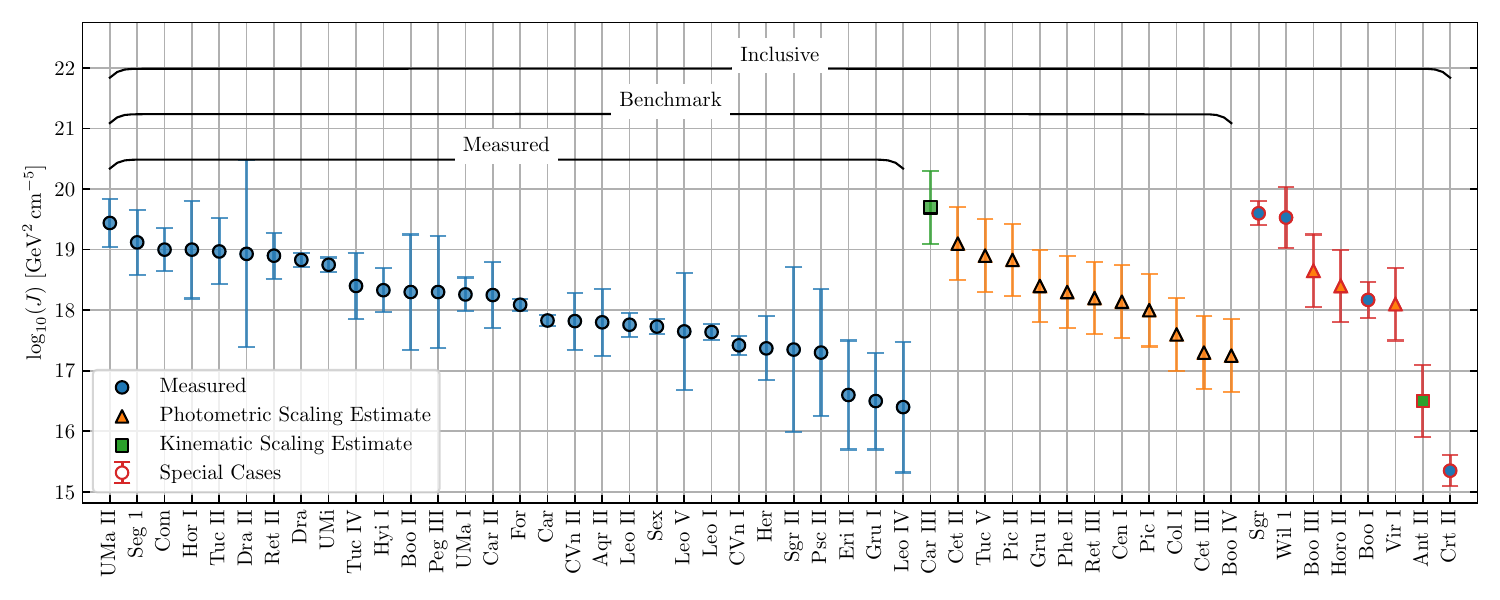}
            \caption{$J$-factors and uncertainties for all \inclusivesize{} dSphs considered in this work. Marker shapes and colors indicate whether the $J$-factor is determined by measurement (blue circles), the kinematic scaling relation of Equation \ref{eq:kinematic} (green squares), or the photometric scaling relation of Equation \ref{eq:photmetric} (orange triangles). The special cases are outlined in red. Also indicated in this figure are the different subsets into which each dSph falls.} \label{fig:Jfactors}
        \end{figure*}

\section{\lat Data Analysis}\label{sec:fermi}
This analysis uses 14.3 years of \lat data taken between August 4, 2008 and December 1, 2022 in the 500 MeV - 1 TeV energy range and selecting photons from the P8R3\_SOURCE\_V3 class. The analysis is performed using \texttt{fermipy} (v1.2) \cite{fermipy}, which utilizes the underlying \texttt{Fermitools} (v2.2.0). The gamma-ray data analysis procedure is similar to previous \Fermi{}-LAT dSphs studies (e.g. \cite{ackermann2014, ackermann2015,drlicawagner2015_dm, albert2017, diMauro2021}).
The Galactic diffuse emission is modeled using the standard interstellar emission model (\texttt{gll\_iem\_v07.fits}) and the emission from point and extended sources is modeled using the 4FGL-DR3 catalog (\texttt{gll\_psc\_v29.fits}) \cite{fgl, dr3}. We perform the analysis in eight logarithmically spaced energy bins per decade with a spatial pixel size of $0.08^{\circ}$. To reduce contamination from photons generated from the Earth’s limb, we use a maximum zenith angle of $100^{\circ}$. Data from the four \Fermi{}-LAT point spread function (PSF) event types are combined in a joint likelihood fit, and the corresponding isotropic spectrum for each PSF type are used (i.e., iso\_P8R3\_SOURCE\_V3\_PSF{\it \{i\}}\_v1, for {\it i} ranging from 0 to 3). For each target coordinate, we define a $10^{\circ}\times 10^{\circ}$ region of interest (ROI), and in order to account for photon leakage from sources outside of the ROI due to the PSF of the detector, the model includes all 4FGL-DR3 sources within a $15^{\circ}\times 15^{\circ}$ region. The energy dispersion correction (edisp\_bins=-1) is enabled for all sources except the isotropic component. The spectral parameters of the Galactic diffuse component (index and normalization) and the normalization of the isotropic component are left free to vary as are the normalizations of all 4FGL sources with a test statistic (TS) value of TS $\geq$ 25 that are within $5^{\circ}$ of the ROI center, and sources with TS $\geq$ 500 and within $7^{\circ}$. The TS is defined as TS $ = 2\log(\mathcal{L}/\mathcal{L}_0))$, where $\mathcal{L}_0$ and $\mathcal{L}$ are the likelihood for the null (i.e., no source present) and alternative hypotheses. The target source is initially modeled as a point source with power law energy spectrum and with the normalization and index free to vary. While we assume a point-source model as has been done in some of the previous dSph analyses \cite{albert2017, diMauro2021}, we also note that some of the dSphs have extensions that could be resolved by \Fermi{}-LAT. Studies of the effects of the dSph extensions on DM limits have been performed \cite{ackermann2014,dimauro2022}. In particular, \cite{dimauro2022} found that the extension of the dSphs could weaken limits on the DM parameter space by a factor of $1.5-1.8$ for a combined dSph analysis. These effects must be considered when discussing constraints derived in this analysis, e.g. regarding  potential tensions with the DM interpretations of the GCE. Additional sources in the ROI are found using \texttt{fermipy}’s \texttt{find\_sources()} function and are included in the model if they have TS $>16$.
In some instances, different ROIs overlap significantly due to small separation between dSph coordinates. To investigate the impact of this we consider Carina II and Carina III, which are the two targets in our sample with the smallest separation of $\sim 0.3^{\circ}$. We fit the ROI with the normalizations of the two sources and the Galactic diffuse free to vary and find the correlation between the normalizations of Carina II and Carina III to be -0.16, indicating they are not highly correlated.

After the analysis procedure described above, the spectral energy distribution (SED) for each dSph is calculated by performing a fit in each energy bin using the \texttt{fermipy} \texttt{sed()} function. The fit in each energy bin assumes a power-law index of 2, and the diffuse background normalizations are left free to vary.
The calculation of the SED yields a likelihood profile,  $\mathcal{L}(d\Phi_{\gamma}/dE, E)$, as a function of energy $E$ and differential gamma-ray flux,  $d\Phi_{\gamma}/dE$. Using the likelihood profile in flux--energy space, a profile can be computed for any given spectral model. To compute the likelihood as a function of DM mass and cross section, we perform the conversion:

\begin{equation}
\mathcal{L}\left(\left<\sigma v \right>, M_{\chi}\right) = \sum_{E_i} \mathcal{L}\left[\frac{d\Phi_{\chi}}{dE}\left(\left<\sigma v \right>, M_{\chi}, E_i\right), E_{i}\right],
\end{equation}
where the sum is performed over the individual energy bins. $d\Phi_{\chi}/{dE}$ is the differential flux for the DM model under consideration where the DM annihilation spectrum ($dN/dE$ of Equation \ref{eq:DMflux}) is provided by PPPC4DM\footnote{\url{http://www.marcocirelli.net/PPPC4DMID.html}} \cite{pppc4dm}. The TS value for a given mass and cross-section pair is given by 

\begin{equation}
TS(\left<\sigma v \right>, M_{\chi})  = 2\left[\frac{\mathcal{L}\left(\left<\sigma v \right>, M_{\chi}\right)}{\mathcal{L}_0}\right],
\end{equation}
where $\mathcal{L}_0$ is the null likelihood (i.e., no source present). We construct TS profiles over the mass range $1\leq M_{\chi}/\mathrm{GeV} \leq 10^4$ and cross-sections $10^{-28}\leq \left<\sigma v\right>/(\mathrm{cm}^{3} \mathrm{s}^{-1})\leq 10^{-22}$, which covers the relevant range of parameter space for GeV-TeV scale thermal relic WIMP DM as well as the constraining capability of \Fermi{}-LAT observations (see e.g. \cite{charles2016}). 
When converting the Poissonian likelihood in flux--energy space to $\left<\sigma v\right> - M_{\chi}$, we also account for the uncertainty in the $J$-factor by multiplying the \Fermi{}-LAT likelihood function with a $J$-factor likelihood function, $\mathcal{L}_J$. 
The $J$-factor likelihood function takes the form of a Gaussian in $\log J$ with width $\sigma_J$,
\begin{eqnarray}
\mathcal{L}_{J}(J) = &&\frac{1}{\ln(10)\sqrt{2\pi \sigma_J} J_{obs}} \nonumber\\
&&\times\exp\left[-\left(\frac{\log_{10}(J)-\log_{10}(J_{obs})}{\sqrt{2}\sigma_J}\right)^2\right],
\end{eqnarray}
where $J_{obs}$ is the $J$-factor value determined either by direct measurement or estimated from scaling relations (cf. \ref{sec:Jfactors}). This is the same $J$-factor likelihood that has been adopted in some previous DM studies using \lat data \cite{ackermann2015, albert2017, diMauro2021}. Once the likelihood profile in $\left<\sigma v\right> - M_{\chi}$ space for each individual dSph is obtained, we sum the individual profiles to obtain the likelihood for the full sample, i.e., the ``combined'' TS profile. 

To quantify the detection significance, we run an analysis on ``blank field'' regions of the sky in which there are no known or potential gamma-ray sources based on the spatial coincidence with the \Fermi{}-LAT and multiwavelength catalogs. The blank-field analysis is used in order to calibrate the null TS distribution and closely follows the procedure of previous \Fermi{}-LAT studies \cite{ackermann2014,ackermann2015,albert2017}. This enables us to account for background effects due to undetected sources in the \Fermi{}-LAT data and imperfect modeling of the diffuse background
emission. The blank-field regions are randomly selected at high Galactic latitudes $(|b|>15^{\circ})$ consistent with the distribution of our dSph sample and are subjected to the same selection criteria applied to the dSph targets: i.e., the central coordinates must not fall within the 95\% radius of 4FGL-DR3 sources and they must be separated by $>0.1^{\circ}$ from any sources in the BZCat, CRATES, and WIBRaLS catalogs (see Section \ref{sec:sample}) in order to minimize potential impacts to unresolved gamma-ray populations \cite{carlson2014}. In total, we analyze 1,000 blank fields. Sets of combined blank fields can then be obtained by selecting randomly and without replacement from the pool of 1,000 a set of analyzed fields of an equal number to the sample under consideration (e.g., \benchmarksize{} for the Benchmark sample). The likelihoods are then added together to calculate the TS value for the combined blank fields. This process is then repeated for 10,000 iterations to construct a sample of combined blank fields. The likelihood profiles for the dSphs and the set of high-Galactic-latitude blank sky locations in both flux--energy space as well as $\left<\sigma v\right>-M_{\chi}$ space are available for download.\footnote{\url{https://figshare.com/articles/dataset/24058650/1}}

In Figure \ref{fig:TS_hists}, we show the distribution of maximum TS values of each combined blank field set across the full mass and cross section range for all annihilation channels considered (i.e., $b\bar{b}$ and $\tau^+\tau^-$). Note that the TS distribution is specific to the sample size (i.e., Measured, Benchmark, and Inclusive sets). These TS distributions are utilized as the null distribution when determining global $p$-values and significances of the combined dSph samples, and to account for the trials factor due to testing different annihilation channels and mass values. 
In an idealized scenario, the TS distribution of the null hypothesis (i.e., the blank-fields) should follow a Poissonian distribution and the asymptotic behavior of Chernoff’s theorem \cite{chernoff} should hold, yielding a distribution compatible with a $\chi^2/2$ distribution for two degrees of freedom in mass and cross-section. However, as can be seen in Figure \ref{fig:TS_hists} there is significant deviation of the observed TS values from the $\chi^2/2$ distribution. This has consistently been observed in previous works \cite{ackermann2014, ackermann2015,dimauro2023_cluster}, and is typically attributed to mismodeled diffuse backgrounds and/or the presence of unresolved gamma-ray sources \cite{carlson2014}. For the individual fields we observe similar deviation from the $\chi^2/2$ distribution as in previous studies, and note that for the combined blank fields the TS distribution varies based on the sample size and $J$-factors used. Given the deviation from statistical expectations, it is critical to use the blank-field distribution in order to calibrate the significance of any observed signal from the dSphs.

\section{Results}\label{sec:results}
\subsection{Individual dSph Analysis}\label{sec:individ_analysis}
In this study, we consider annihilation channels into $b\bar{b}$ and $\tau^+\tau^-$ final states.  These bracket most of the supersymmetry model space \cite{cotta2012} and are typically taken to be representative annihilation channels. Furthermore, annihilation through these channels has been shown to be compatible with the GCE – particularly in the case of $b\bar{b}$ final states. The $\tau^+\tau^-$ channel is also of particular interest in DM studies performed by atmospheric or water Cherenkov telescopes such as H.E.S.S., MAGIC, VERITAS, HAWC, or the upcoming Cherenkov Telescope Array, as they are particularly sensitive to the harder photon spectra produced (see e.g., \cite{HESS,archambault2017,albert2018,strigari2018, acharyya2021,acciari2022}). Electroweak corrections for the $\tau^+\tau^-$ channel are accounted for in the annihilation spectrum as implemented in the PPPC4DM. The electroweak corrections for the quark channels are negligible for the energy range considered in this work \cite{ciafaloni2011}, and we do not account for them in the $b\bar{b}$ channel.

In the individual analysis of the dSphs no individual targets reach the typical TS benchmark of TS $>25$ used by the {\it Fermi}-LAT Collaboration to designate a detection as significant. In order to quantify the significance of the signal from individual dSphs we compare these results with the analysis of the individual blank fields. In Figure \ref{fig:TSvm_individual}, we show the maximum TS over all cross-section values as a function of the DM mass for the individual dSphs, compared with the 97.5\% and 84\% containment bands of the individual blank fields. Seven of the dSphs exhibit excesses with local significance $\geq 2\,\sigma$ for both annihilation channels (Bo\"{o}tes III, Hercules, and Ursa Minor all have local significance $\sim 2.1\, \sigma$ for the $\tau^+\tau^-$ channel only). Of these seven dSphs, Tucana II, Willman 1, Reticulum II, Horologium II and Bootes I were also reported as among the highest detection significance sources in the analysis of \cite{diMauro2021}. Interestingly, the individual fields with higher TS tend to also have relatively large $J$-factors. In fact, Carina III and Willman 1 have the two largest $J$-factors of all the dSphs (neglecting upper limits), and Tucana II and Reticulum II have the ninth and twelfth largest. Figure \ref{fig:flux_ul} shows the flux upper limits for the dSphs compared to their $J$-factor values, as well as the bands for the 68\% and 95\% containment bands for the individual blank-fields. We highlight in this figure the dSphs with local significances $\geq 2\,\sigma$ in both channels, which tend toward higher flux upper limits and $J$-factors. However, we note that these dSphs do not always have their local significances peak at consistent DM masses, which would be expected in the case that each local excess could be attributed to a purely DM induced signal.

The largest excess is for Reticulum II, with a local significance peak in the mass range $290 < M_{\chi}/\mathrm{GeV} < 588$ ($27 < M_{\chi}/\mathrm{GeV} < 70$)  for the $b\bar{b}$ ($\tau^+\tau^-$) annihilation channel, and at these masses the TS exceeds all of the blank fields. The marginal excess in Reticulum II has been observed in a number of previous studies \cite{drlicawagner2015_dm, diMauro2021, geringerSameth2015_ret2, hooperLinden2015_ret2}. Ref. \cite{diMauro2021} finds a local significance of $2.6-2.8 \,\sigma $ over a similar range of masses as this study for each channel, while \cite{drlicawagner2015_dm, geringerSameth2015_ret2, hooperLinden2015_ret2} found local excesses ranging from $2.4-3.2\,\sigma$. These latter works found masses typically lower than observed in this analysis, though still compatible in the case of \cite{drlicawagner2015_dm, geringerSameth2015_ret2}, while Ref \cite{hooperLinden2015_ret2} only reported results for testing a GCE motivated DM model of a $49$ GeV WIMP annihilating through the $b\bar{b}$ channel. We also note that these works were performed using older \Fermi{} catalogs (3FGL), older diffuse models, and about half the \Fermi{} exposure used in this analysis.

The global significance of the individual dSphs is determined relative to the distribution of the blank-field analysis. Specifically, the distribution of TS values of the individual blank fields over all masses and cross sections is treated as the null distribution. We then obtain a $p$-value from this distribution at the TS value of the local significance peak for the dSphs, which is converted to a significance assuming a one-tailed standard normal distribution. In this way, our global significance includes a trials factor for the fact that we have searched over a range of masses and annihilation channels. From these we obtain for the $b\bar{b}$ channel $p_{global} = 0.016$ ($2.1 \,\sigma$) for Reticulum II. We also note Hydrus I and Horologium II, which have peak TS values of 10.3 and 10.8, and corresponding global significances of $\sim 2.2\,\sigma$, though their maximum local significances are lower than that of Reticulum II. This is due to the broader TS distribution of the blank fields at the lower masses that are best fit to Hydrus I and Horologium II. Additionally, Horologium II was flagged (see Section \ref{sec:sample}) due to spatial coincidence ($\sim 0.02^{\circ}$ separation) with a blazar candidate in the WIBRaLS catalog, which may be a contaminating source of gamma-ray emission for this target. We note that the $J$-factor and uncertainty do not affect the individual detection significance; however, they do affect the DM interpretation and the significance of the combined results. Table \ref{tab:indiv_results} summarizes the results for the individual dSphs, including the DM mass with the highest local significance and the corresponding local and global $p$-values and significances for the dSphs with a local significance $\geq 2\,\sigma$ for both $b\bar{b}$ and $\tau^+\tau^-$ annihilation channels. We also show in Table \ref{tab:indiv_results} the quantity $p_{sample}$, which is the $p$-value after applying a trials factor penalty due to observing 50 different dSphs. This is calculated as $p_{sample}=1-(1-p_{global})^{N_{dSphs}}$ where $N_{dSphs} =50$ is the total number of dSphs analyzed \cite{gustaffson2013,albert2017}. Given the relatively large number of dSphs, the application of this additional trials factor greatly reduces the overall significance for each dSph determined by $p_{sample}$.

\begin{table*}[tbph!]
\centering
\def\arraystretch{1.55}
\setlength{\tabcolsep}{4pt}

\begin{tabular}{lcccc}

\hline\hline

 & $M_{\chi}$ &  $p_{\mathrm{local}}$ & $p_{\mathrm{global}}$ & $p_{\mathrm{sample}}$ 
\\
 & [GeV]                    &                     & & \\
\hline
 & \multicolumn{4}{c}{$b\bar{b}$} \\
 \hline

Bo\"{o}tes II & 228.5 & $1.3\times 10^{-2}$ ($2.2 \, \sigma$) & $8.1\times 10^{-2}$ ($1.4 \,\sigma$)& 0.97 $(-1.9\,\sigma)$ \\
Carina III & 366.5 & $2.1\times 10^{-2}$ ($2.0 \, \sigma$) & $1.1\times 10^{-1}$ ($1.2 \,\sigma$)& 0.99 $(-2.5\,\sigma)$ \\
Horologium II & 8.4 & $5.0\times 10^{-3}$ ($2.6 \, \sigma$) & $1.3\times 10^{-2}$ ($2.2 \,\sigma$)& 0.50 $(-0.0\,\sigma)$ \\
Hydrus I & 8.4 & $4.0\times 10^{-3}$ ($2.7 \, \sigma$) & $1.2\times 10^{-2}$ ($2.2 \,\sigma$)& 0.50 $(-0.0\,\sigma)$ \\
Reticulum II & 289.4 & $<10^{-3}$ ($>3 \, \sigma$) & $1.6\times 10^{-2}$ ($2.1 \,\sigma$)& 0.57 $(-0.2\,\sigma)$ \\
Tucana II & 13.4 & $9.9\times 10^{-3}$ ($2.3 \, \sigma$) & $2.8\times 10^{-2}$ ($1.9 \,\sigma$)& 0.73 $(-0.6\,\sigma)$ \\
Willman 1 & 228.5 & $2.0\times 10^{-3}$ ($2.9 \, \sigma$) & $3.3\times 10^{-2}$ ($1.8 \,\sigma$)& 0.76 $(-0.7\,\sigma)$ \\
\hline
&\multicolumn{4}{c}{$\tau^+\tau^-$}\\
 \hline
Bo\"{o}tes II & 13.4 & $1.6\times 10^{-2}$ ($2.1 \, \sigma$) & $5.9\times 10^{-2}$ ($1.6 \,\sigma$)& 0.97 $(-1.9\,\sigma)$ \\
Carina III & 43.8 & $1.2\times 10^{-2}$ ($2.3 \, \sigma$) & $8.3\times 10^{-2}$ ($1.4 \,\sigma$) &0.99 $(-2.5\,\sigma)$ \\
Horologium II & 2.6 & $< 10^{-3}$ ($>3 \, \sigma$) & $7.9\times 10^{-3}$ ($2.4 \,\sigma$)& 0.36 $(0.4\,\sigma)$ \\
Hydrus I & 2.6 & $4.0\times 10^{-3}$ ($2.7 \, \sigma$) & $1.7\times 10^{-2}$ ($2.1 \,\sigma$) &0.57 $(-0.2\,\sigma)$ \\
Reticulum II & 27.3 & $<10^{-3}$ ($>3 \, \sigma$) & $2.2\times 10^{-2}$ ($2.0 \,\sigma$) &0.68 $(-0.5\,\sigma)$ \\
Tucana II & 5.2 & $6.9\times 10^{-3}$ ($2.5 \, \sigma$) & $1.4\times 10^{-2}$ ($2.2 \,\sigma$) &0.50 $(-0.0\,\sigma)$ \\
Willman 1 & 43.8 & $<10^{-3}$ ($>3 \, \sigma$) & $3.6\times 10^{-2}$ ($1.8 \,\sigma$)& 0.88 $(-1.2\,\sigma)$ \\
\hline\hline
\end{tabular}
\caption{DM mass and local and global $p$ values (and significances) of individual dSphs with local signifcances $\geq 2\, \sigma$ for both annihilation channels.}\label{tab:indiv_results}
\end{table*}

\begin{figure*}
        \centering
        \includegraphics[width=0.9\linewidth]{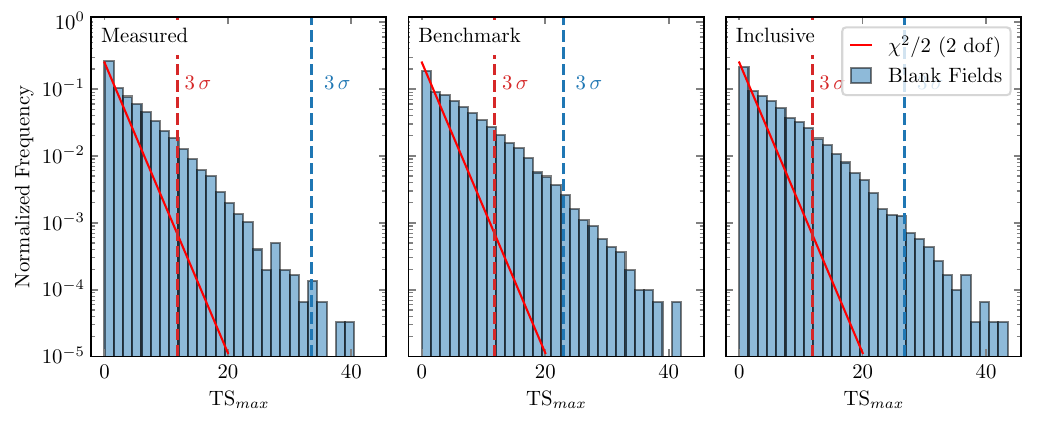}
        \caption{TS distribution for combined blank fields corresponding to each of our sub samples. 
        For reference, we show the distribution of $\chi^2/2$ with two degrees of freedom, as well as vertical lines comparing the TS value corresponding to $3\,\sigma$ for the Chernoff's theorem approach (red-dashed) vs the empirical method used in this analysis (blue-dashed) for a given subsample.}\label{fig:TS_hists}
\end{figure*}

\begin{figure*}
     \centering
    \includegraphics[width=3.5in]{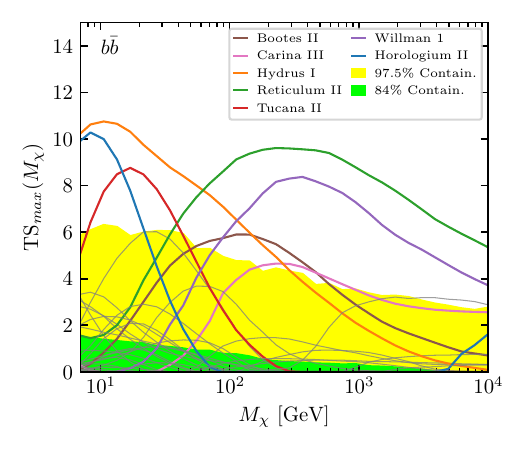}\hfill
    \includegraphics[width=3.5in]{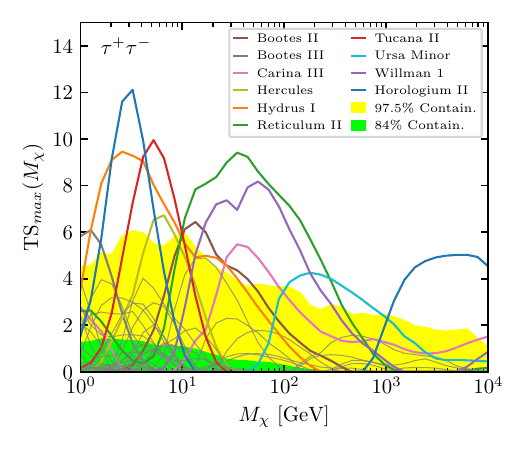}
    \caption{Maximum TS as a function of $M_{\chi}$ over all cross-section values for the individual dSphs in comparison with the individual blank fields. The dSphs with a local significance $\geq 2\,\sigma$ are shown as colored lines. The yellow and green bands show the 97.5\% and 84\% containment regions for the individual blank fields for the $b\bar{b}$ ({\it left}) and $\tau^+\tau^-$ ({\it right}) annihilation channels.
    \label{fig:TSvm_individual}
    }
\end{figure*}

\begin{figure*}
     \centering
    \includegraphics[width=3.5in]{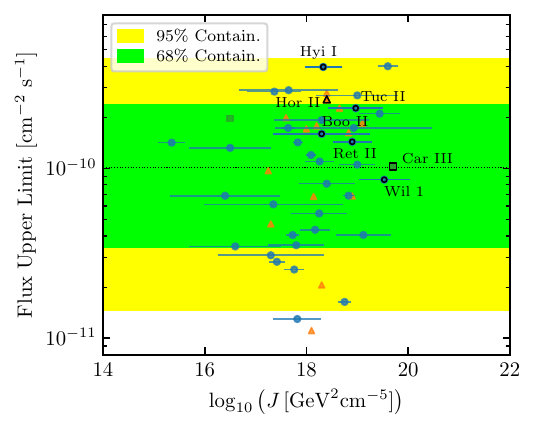}\hfill
    \includegraphics[width=3.5in]{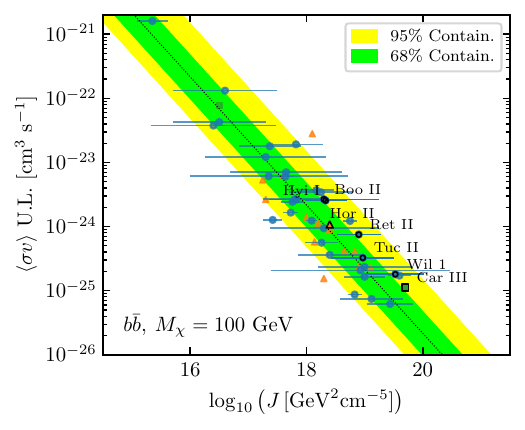}
    \caption{Flux ({\it left}) and cross section ({\it right}) upper limits vs $J$-factor for the individual dSphs. The upper limits on cross-section are calculated assuming $M_{\chi}=100$ GeV and $b\bar{b}$ annihilation channel. The dSphs with local significances $\geq 2\,\sigma$ for both annihilation channels are are outlined in black and labeled. Colors and shapes are defined the same as Figure \ref{fig:Jfactors}, except that Special cases are not highlighted here for visual clarity. Error bars are not shown for the dSphs without measured $J$-factors where an uncertainty of 0.6 dex is assumed.\label{fig:flux_ul}
    }
\end{figure*}

\begin{table*}[tbph!]
\centering
\def\arraystretch{1.55}
\setlength{\tabcolsep}{4pt}

\begin{tabular}{lccc|ccc}

\hline\hline
 & \multicolumn{3}{c}{$b\bar{b}$} & \multicolumn{3}{c}{$\tau^+\tau^-$}\\
 \cline{2-4}  \cline{5-7}
 & $M_{\chi}$ &  $p_{\mathrm{local}}$ & $p_{\mathrm{global}}$  & $M_{\chi}$ &  $p_{\mathrm{local}}$ & $p_{\mathrm{global}}$
\\
 & [GeV]                    &                     &  & [GeV]                    &                     &
\\
\hline
Measured & \measplocmaxMass{} & \measplocmax{} (\meassloc{}) & \measpglobal{} (\meassglobal{}) & \measplocmaxMassTau{} & \measplocmaxTau{} (\measslocTau{}) & \measpglobalTau{} (\meassglobalTau{})
\\
Benchmark & \benchplocmaxMass{} &  \benchplocmax{} (\benchsloc{})&  \benchpglobal{} (\benchsglobal{}) & \benchplocmaxMassTau{} &  \benchplocmaxTau{} (\benchslocTau{})&  \benchpglobalTau{} (\benchsglobalTau{})
\\
Inclusive & \incplocmaxMass{} & \incplocmax{} (\incsloc{}) & \incpglobal{} (\incsglobal{}) & \incplocmaxMassTau{} & \incplocmaxTau{} (\incslocTau{}) & \incpglobalTau{} (\incsglobalTau{})
\\
\hline\hline
\end{tabular}
\caption{DM mass and local and global $p$ values (and significance) for each sample at the local significance peak.}\label{tab:results}
\end{table*}

\subsection{Combined dSph Analysis}

In the combined likelihood analysis of the dSphs, we sum the likelihood profiles of all dSphs in a given sample to determine the combined signal. The maximal TS values of each sample are not found to be significant when compared with the global TS distributions. The Measured sample which contains only dSphs with measured $J$-factors does not exhibit any local excesses over the mass range considered for either annihilation channel, with local significance $<2\,\sigma$ throughout. However, in the Benchmark and Inclusive samples we find that the dSphs exhibit marginal excesses relative to the background over some region of the $\left<\sigma v\right>-M_{\chi}$ space for both $b\bar{b}$ and $\tau^+\tau^-$ annihilation channels. In the top panel of Figure~\ref{fig:TSvM_sample}, we show the TS profile as a function of the mass for the combined samples. The shaded regions represent the 97.5\% containment of the combined blank fields. We see that the Benchmark and Inclusive subsamples reach or exceed the background level above the 97.5\% containment over some portion of the mass range whereas the Measured sample does not. The disparity between the Measured sample and the other two samples can be attributed to the absence of some high TS dSphs from the Measured sample. For instance, the difference between the Measured and Benchmark sample at $M_{\chi}\sim 200$ GeV for the $b\bar{b}$ channel is almost entirely due to the exclusion of Carina III from the Measured sample. To directly quantify the excesses relative to background for each subsample, we show the confidence intervals as a function of the DM mass in the lower panel of Figure~\ref{fig:TSvM_sample}. The horizontal lines indicate the corresponding significance levels assuming a one-tailed standard normal distribution. Since we are comparing the TS of the dSph samples with the TS distribution of the combined blank fields {\it at a specified mass,} these values represent the {\it local} significance. In our Benchmark sample we observe an excess relative to the background for the $b\bar{b}$ ($\tau^+\tau^-$) annihilation channel at a local significance of \benchsloc{} (\benchslocTau{}) with a mass of \benchplocmaxMass{} GeV (\benchplocmaxMassTau{} GeV) for the Benchmark sample. While this does not constitute a significant detection, it is an interesting result and motivates future studies with more data and larger sample sizes. For the Inclusive sample, which comprises all confirmed and probable dSphs including the ``Special'' cases, the local significance peaks at a mass of \consplocmaxMass{}GeV (\consplocmaxMassTau{}GeV) with a local significance of \incsloc{} (\incslocTau{}). However, as noted in Section \ref{sec:sample}, the results from the Inclusive set should be considered with caution. This set includes some sources that may be contaminated by other astrophysical gamma-ray sources due to their spatial coincidence with sources in multiwavelength catalogs (e.g. blazars and blazar candidates). In contrast, the blank field set excludes fields that are spatially coincident with multiwavelength catalog sources, which occurs at rate of about $\sim 2\%$ for the random uniform sampling at $|b|>15^{\circ}$.

While the local excesses above background are interesting, we must also consider the effect of scanning over several mass values and testing multiple annihilation channels -- i.e., the ``look elsewhere'' effect. This is accounted for by comparing the TS at the peak of the local significance of the combined dSph samples with the relevant TS distribution as shown in Figure \ref{fig:TS_hists}, which yields the global $p$-values and significances accounting for the mass and annihilation channel trials. For the $b\bar{b}$ annihilation channel, we find global $p$-values of \measpglobal{} (\meassglobal{}), \benchpglobal{} (\benchsglobal{}), and \incpglobal{} (\incsglobal{}) for the Measured, Benchmark, and Inclusive subsets, respectively. Similar values are found for the $\tau^+\tau^-$ channel. The results for each subsample and annihilation channel are summarized in Table \ref{tab:results}. When accounting for the extra trials due to testing an ensemble of different masses and annihilation channels, the global significances drop to $< 1\,\sigma$. 

We note that in general the peak of the  local significance does not necessarily correspond to the overall peak of the TS, due to the mass-dependent TS profile of both the combined dSphs and blank-fields. For example, in the Benchmark sample there is a peak in the TS profile at $M_{\chi}\sim 10$ GeV for the $b\bar{b}$ annihilation channel. This low-mass peak has a lower local significance than that found at $\sim 180 $ GeV, as can be seen by comparison with the 97.5\% containment band of the combined blank fields in Figure \ref{fig:TSvM_sample}, top. The presence of the low-mass peak is consistent with background within $<2\,\sigma$ local significance and appears to be due in particular to Hydrus I and Tucana II, which exhibit TS peaks at relatively low mass values (see Figure \ref{fig:TSvm_individual}). Similar low-mass TS peaks are also observed in the blank fields. In part, these peaks in the joint likelihood analysis are a product of the uncertainty in the $J$-factor. Increases in the assumed $J$-factor uncertainty tend to amplify these low-mass TS values, while decreased uncertainties tend to mitigate them. This can have noticeable impacts on the global significance, however the effect on the local significance is minor. In the extreme limiting case where the $J$-factor prior is not applied (i.e., no uncertainty), this low-mass peak in the combined sample does not exist. They are also likely influenced by the greater systematic uncertainty at low energies. In Appendix \ref{sec:weight}, we consider the possibility that the low-mass TS peaks are connected to background fluctuations amplified by the $J$-factor uncertainty. To do this, we perform a weighted likelihood analysis following the approach used for the 4FGL catalogs \cite{fgl}.

\begin{figure*}
     \centering
    \includegraphics[width=3.5in]{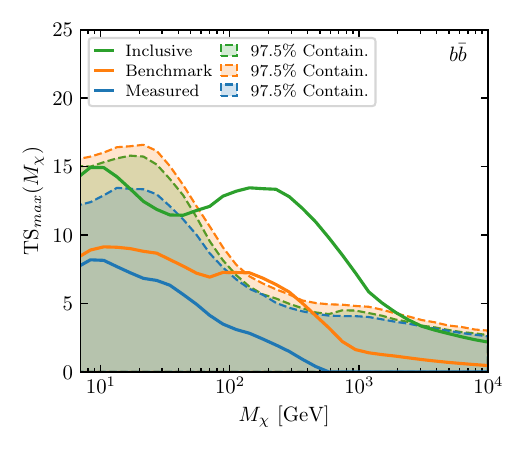}\hfill \includegraphics[width=3.5in]{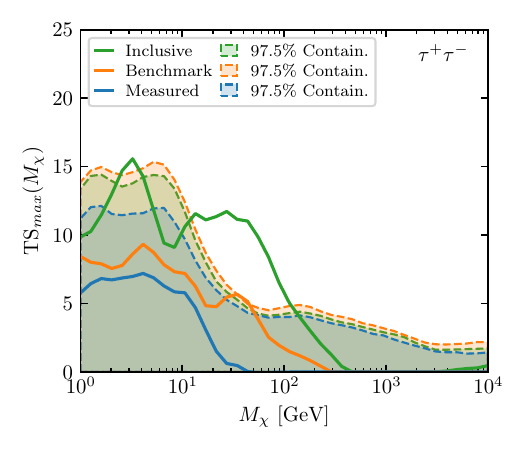}\\
    \includegraphics[width=3.5in]{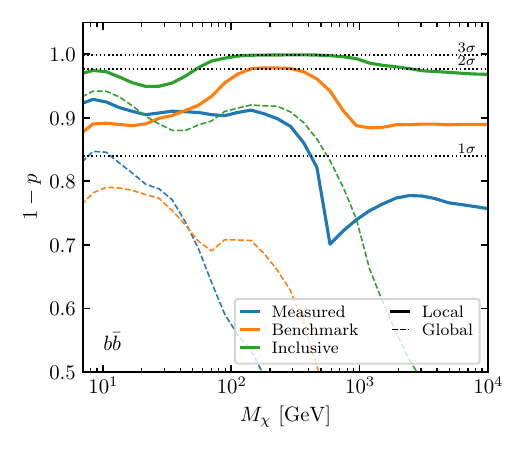}\hfill \includegraphics[width=3.5in]{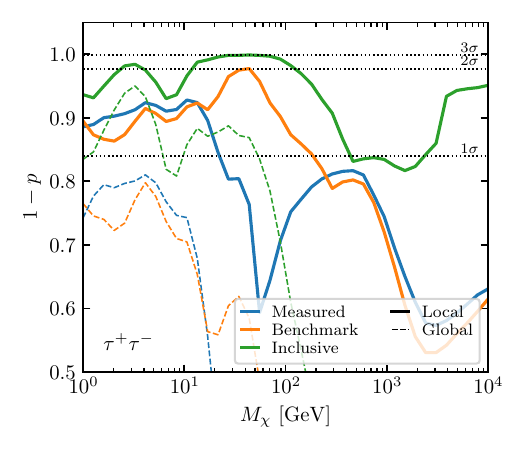}
    \caption{{\bf Top:} Maximum TS as a function of $M_{\chi}$ over all cross-section values for each dSph sample; shaded regions are the 97.5\% containment region for the combined blank fields. {\bf Bottom:} $1-p$ for each sample with respect to the combined blank fields. Solid lines show the local values ($1-p_{local}$) while dashed lines show the global values ($1-p_{global}$). Horizontal dotted lines indicate the 1, 2, and 3 $\sigma$ levels.
    \label{fig:TSvM_sample}
    }
\end{figure*}



\subsection{Dark Matter Constraints}
Given the lack of a significant detection from the sample of dSphs, we derive upper limits on the DM cross-section. The upper limits are obtained using the ``delta-log-likelihood'' approach, where the limit at each mass is set at the cross section for which the TS decreases by 2.71 (i.e. $2\Delta \log\mathcal{L}=-2.71$) \cite{ackermann2014, ackermann2015}.
Figure \ref{fig:uls} shows the constraints derived in this analysis for each annihilation channel compared with previous results. Specifically, the top row compares the upper limits from the Measured sample with the limits from \cite{ackermann2015}, which similary used only dSphs with measured $J$-factors. The lower panels show the upper limits derived from the Benchmark sample in comparison with previous constraints from \cite{albert2017} and \cite{diMauro2021}.
The limit from the Inclusive sample is qualitatively similar and provides effectively the same constraining power, so for visual clarity we do not include it in the plot. Also shown in each panel are a selection of best-fit models for the GCE from \cite{GordonMacias2013, calore2015, daylan2016, abazajian2016,karwin2017_GCE, diMauro2021}. The constraints from the Measured sample are comparable to those of \cite{ackermann2015} for DM masses $\lesssim 1-2$ TeV and $\lesssim 100-200$ GeV for the $b\bar{b}$ and $\tau^+\tau^-$ annihilation channels respectively, after which there is slight improvement from the Measured sample. For the Benchmark sample, the limits slightly improve upon those presented in \cite{albert2017} and \cite{diMauro2021} at masses between $20-100$ GeV for the $b\bar{b}$ channel and $<1$ TeV for the $\tau^+\tau^-$ channel. Otherwise they are very similar, with the exception of being slightly weaker than the limits from \cite{diMauro2021} at $M_{\chi}>1$ TeV in the  $\tau^+\tau^-$ channel.
This analysis has further constrained the DM parameters space relative to previous results and is similarly in tension with several models proposed as possible explanations of the GCE. 
Despite the tension between the dSphs limits and the GCE models, these updated constraints are not stringent enough to confidently rule out the GCE models. In particular, when considering the uncertainties in the DM density profile of the GC and the systematics of the GCE spectrum, the GCE models are still compatible with the dSphs limits.

\begin{figure*}
    \includegraphics[width=3.5in]{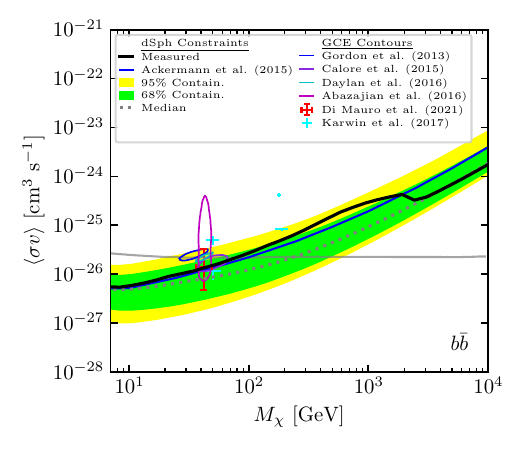}\hfill
    \includegraphics[width=3.5in]{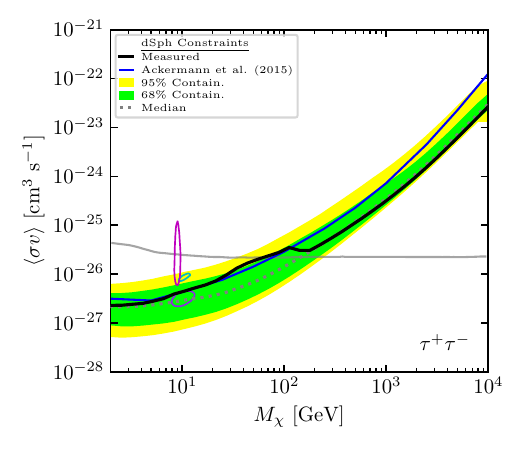}\\
    \includegraphics[width=3.5in]{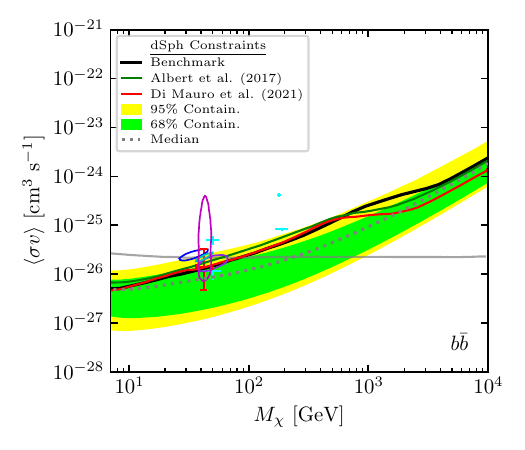}\hfill
    \includegraphics[width=3.5in]{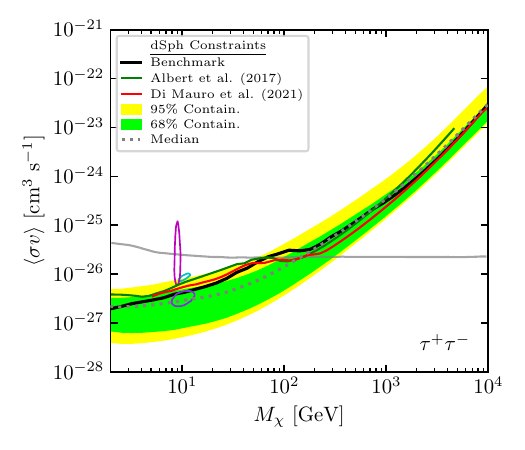}
    \caption{Constraints on the $b\bar{b}$ (left) and $\tau^+\tau^-$ (right) annihilation channels derived in this work compared to previous dSph constraints as well as GCE models from \cite{GordonMacias2013, calore2015, daylan2016, abazajian2016, karwin2017_GCE,diMauro2021}. The contour definitions in the top left panel legend apply to each figure. The yellow and green bands show the 95\% and 68\% containment regions respectively of the upper limits derived from the combined blank fields and the gray line is the thermal relic cross-section from \cite{steigman}. {\bf Top:} Comparison of the upper limits from the Measured sample with previous results from the dSphs analysis of \cite{ackermann2015}. {\bf Bottom:} Comparison of the upper limits from the Benchmark sample with previous results from \cite{albert2017} and \cite{diMauro2021}.
    \label{fig:uls}}
\end{figure*}

\subsection{Projections for Future Dark Matter Searches with the \Fermi{}-LAT}

Here we present a brief discussion of sensitivity projections for future dSph studies in light of the results presented in this manuscript. Overall, the total sensitivity improvements of future dSph studies are impacted by additional data obtained from \Fermi{}-LAT observations, potential increases in the target sample from optical sky surveys (e.g., using DECam and {\it Rubin}) as well as improved measurements of the astrophysical $J$-factor for currently known and/or newly discovered satellites achieved with improved stellar kinematic studies. 

Ref. \cite{charles2016} provides a general discussion of the impacts these factors can have on projected sensitivity. For additional observing time, sensitivity improvements scale as $1/\sqrt{t}$ for background limited DM searches or $1/t$ for signal limited searches. As detailed in \cite{charles2016}, whether the dSph analysis falls under the background-limited or signal-limited regime depends to some extent on the DM annihilation spectrum (and by proxy the DM mass and annihilation channel), and is better described by a mixture of the two regimes. However for a conservative estimate, we adopt the $1/\sqrt{t}$ scaling. Accounting for a potential increase in the number of dSphs analyzed in the sample is a more complicated and uncertain process. For the purposes of this estimate, we assume that the increase in sensitivity scales with increases of the sample in the same manner determined in \cite{charles2016}. However, the actual improvement in the sensitivity would be dependent on the $J$-factors and distances of the additional targets, which are unknown and may differ from the values of the known dSphs. 

We show the improvement in sensitivity of our analysis compared to the previous analyses of \cite{ackermann2015, albert2017} in the left panel of Figure \ref{fig:sensitivity}. This is illustrated by comparing the median of the blank-fields from those works with the median of the blank-fields from the Measured and Benchmark samples of our analysis. We also show the predicted sensitivities of \cite{charles2016} based on 15 total years of \Fermi{}-LAT exposure and increased dSph sample sizes. It is evident from this figure that the additional exposure and larger dSph sample in our analysis has yielded improved sensitivity and that the improvement in sensitivity is generally compatible with expectations from \cite{charles2016}. For masses $\gtrsim 200 $ GeV, the sensitivity of the blank fields corresponding to the Measured sample is nearly identical to the predictions for a 30 dSph sample and 15 years of exposure presented in \cite{charles2016}, although the lower mass sensitivity has not improved as projected. The sensitivity limits provided by the blank fields corresponding to the Benchmark sample are similar above $\gtrsim 200$ GeV to the projections from \cite{charles2016} for a sample size of 60 dSphs. 

The right panel of Figure \ref{fig:sensitivity} shows the projected improvement in the local signal significance assuming a $b\bar{b}$ annihilation channel for the Benchmark sample with additional {\it Fermi}-LAT exposure and various increases in the number of dSphs. Signal improvements are projected assuming that the significance improves with time as $s \propto \sqrt{t}$, and scales with additional dSphs as presented in Ref. \cite{charles2016} for the $b\bar{b}$ channel. As noted above, the observed sensitivity improvement in this analysis is consistent with the expectations from \cite{charles2016} for higher masses, but deviates substantially at lower masses. Therefore, while adopting the sensitivity scaling of \cite{charles2016} is well-supported for our peak TS mass range ($150-230$ GeV), this assumption can not be reliably applied at lower masses. For our Benchmark sample of dSphs, we found a \benchmarkSig{} local significance relative to background (Sec \ref{sec:results}), corresponding to \benchsglobal{} global significance when including the trials factor for testing different masses and annihilation channels. From Figure \ref{fig:sensitivity}, we see that in order to obtain a $5\,\sigma$ local significance detection within an additional 10 years of {\it Fermi}-LAT observations would require an increase in the known dSph population of about 65 dSphs (or $\sim 35$ for $4\,\sigma$). In terms of the global significance, increases of 65 and 35 dSphs with an additional 10 years of \Fermi{}-LAT data would yield global significances of $1.2\,\sigma$ and $1\,\sigma$ respectively. These increases in the dSph sample are possible given the results of ongoing dSph searches (e.g., \cite{cerny2021,cerny2023,cerny2023b, smith2022}), as well as the upcoming {\it Rubin} Legacy Survey of Space and Time \cite{ivezic2019}. Thus, combining additional {\it Fermi}-LAT observations with the anticipated increases in the known dSph population can offer the opportunity to potentially achieve a statistically significant detection of these systems within a feasible timeline. 

\begin{figure*}
    \includegraphics[width=3.5in]{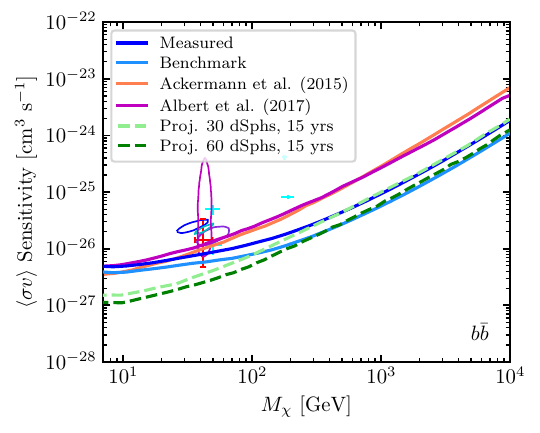}\hfill
    \includegraphics[width=3.3in]{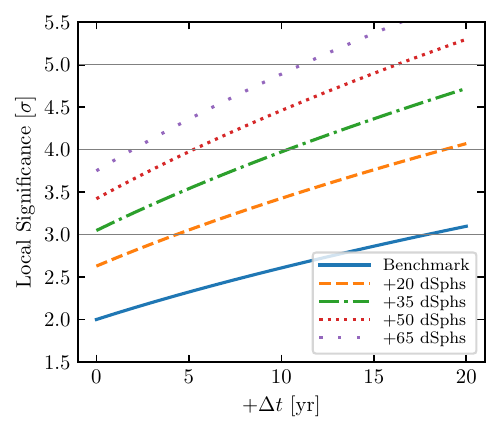}
    \caption{{\bf Left:} Comparison of sensitivity between the Measured (\measuredsize{} dSphs) and Benchmark (\benchmarksize{} dSphs) samples of this analysis with the sensitivity limits from \cite{ackermann2015,albert2017} as determined by the median of the blank-fields limits. Light and dark green dashed curves are the predicted sensitivity limits from \cite{charles2016} for 15 total years of \Fermi{}-LAT exposure and different dSph sample sizes. The contours and data points are the GCE models as defined in Figure \ref{fig:uls}.{\bf Right:} Projected local signal significance relative to background for the Benchmark sample given further {\it Fermi}-LAT exposure and with additional dSphs in the sample.
    \label{fig:sensitivity}}
\end{figure*}

\section{Conclusion}\label{sec:conclusion}
In this work, we have analyzed a sample of Milky Way satellite dSphs with 14.3 years of \lat data to search for a gamma-ray signal from annihilating DM. No significant (e.g., TS $>25$; $>5\,\sigma$) detection is made for any dSph individually or through the joint analysis of the dSph subsamples. However, while we do not find any significant detections from the individual dSphs, seven dSphs have excesses relative to background at a local significance level of $\geq 2\,\sigma$ for both annihilation channels, although accounting for trials factors yields smaller global significances of roughly  $\lesssim 2\,\sigma$.  The most significant local excess is for Reticulum II with a local significance $>3\,\sigma$ ($\sim 2\,\sigma$ global) at masses of $290 < M_{\chi}/\mathrm{GeV} < 588$ for the $b\bar{b}$ annihilation channel  and $27 < M_{\chi}/\mathrm{GeV} < 70$ for the $\tau^+\tau^-$ annihilation channel. Other dSphs with the highest local significance include Willman 1, Horologium II and Hydrus 1. A joint likelihood analysis is used to determine the DM annihilation signal from the full sample. Noting that there is uncertainty in the DM content of some of the targets, we consider three subsets of the full sample. Qualitatively, limits from each sample are fairly similar. 
We place updated constraints on the $M_{\chi}-\left<\sigma v\right>$ parameter space that provide mostly consistent, though slightly tighter, restrictions relative to other recent studies of dSphs \cite{ackermann2015, albert2017, diMauro2021}. Thus, the \Fermi{}-LAT results remain among the most constraining probes of WIMP DM to date. Furthermore, as in some previous works, the new constraints are in tension with DM models that have been proposed as potential origins for the GCE \cite{GordonMacias2013, calore2015, daylan2016, abazajian2016, dimauro2021_GCE}. 
 
While a significant signal is not found in this work, we observe a \benchmarkSigEst{} local (\benchsglobal{} global) excess relative to the the background estimated from blank-field regions at a DM mass of $M_{\chi}=$ \benchmarkMassrange{} GeV for $b\bar{b}$ annihilation and \benchmarkSigEstTau{} (\benchsglobalTau{} global) at $M_{\chi}$ =\benchmarkMassrangeTau{} GeV for $\tau^+\tau^-$ in the Benchmark sample. Some analyses have indicated that a $ 100-200$ GeV DM particle annihilating to quarks may be a viable candidate to explain the GCE \cite{karwin2017_GCE}, which would be consistent with the observed excess in the dSphs. However, the global significance of this excess is $<1\,\sigma$ once the trials factor from scanning multiple masses and channels in our search is accounted for. The public release of the dSph likelihood profiles will enable further testing of various DM models. Additionally, subsequent studies incorporating other relevant properties of the dSphs such as their spatial extension (e.g., \cite{dimauro2022}) may also provide further insight into the possible DM annihilation from the dSphs. 

The improved sensitivity of our search relative to previous searches is an encouraging sign that future studies may improve  upon these results with the addition of newly detected Milky Ways dSphs that may be discovered by optical sky surveys, as well as additional observational data acquired by the {\it Fermi}-LAT. We estimate that the signal significance could approach the $5 \,\sigma$ local ($1.2 \,\sigma$ global) significance level with an additional $ 10-15$ years of \Fermi{}-LAT data and the discovery of 50-65 dSphs, or a $4\,\sigma$ ($1\,\sigma$ global) level in a similar $\sim 10$ year time frame with a more modest increase of $\sim 35$ additional dSphs.

\section*{Acknowledgements}
AM and MA acknowledge funding under NASA contract 80NSSC22K1580. CMK's research was supported by an appointment to the NASA Postdoctoral Program at NASA Goddard Space Flight Center, administered by Oak Ridge Associated Universities under contract with NASA. ADW was partially supported by the NASA Fermi Guest Investigator Program Cycle 9 No. 91201. Clemson University is acknowledged for generous allotment of compute time on the Palmetto cluster. 

The \textit{Fermi}-LAT Collaboration acknowledges generous ongoing support
from a number of agencies and institutes that have supported both the
development and the operation of the LAT as well as scientific data analysis.
These include the National Aeronautics and Space Administration and the
Department of Energy in the United States, the Commissariat \`a l'Energie Atomique
and the Centre National de la Recherche Scientifique / Institut National de Physique
Nucl\'eaire et de Physique des Particules in France, the Agenzia Spaziale Italiana
and the Istituto Nazionale di Fisica Nucleare in Italy, the Ministry of Education,
Culture, Sports, Science and Technology (MEXT), High Energy Accelerator Research
Organization (KEK) and Japan Aerospace Exploration Agency (JAXA) in Japan, and
the K.~A.~Wallenberg Foundation, the Swedish Research Council and the
Swedish National Space Board in Sweden.
 
Additional support for science analysis during the operations phase is gratefully
acknowledged from the Istituto Nazionale di Astrofisica in Italy and the Centre
National d'\'Etudes Spatiales in France. This work performed in part under DOE
Contract DE-AC02-76SF00515.

\appendix
\section{Weighted Likelihood Analysis}\label{sec:weight}

As can be seen in Figures \ref{fig:TSvm_individual}  and \ref{fig:TSvM_sample}, the effect of background fluctuations amplified by the $J$-factor uncertainty can produce relatively large TS at low masses. This may in part be attributable to systematic uncertainties in the diffuse model around the target that are particularly impactful at lower energies (and thus, lower DM masses). In compiling the latest \Fermi{} catalogs (i.e., 4FGL and later) a weighted log-likelihood method is implemented to mitigate these effects. We therefore also implement the weighted likelihood method on our analysis of the dSphs and blank fields in order to investigate the effects of these systematics. Specifically, what contribution to the low-mass TS peaks may be attributable to systematics of the diffuse model. The procedure for this is exactly the same as our standard analysis, however we use the likelihood weighting as applied in the 4FGL catalog (see Appendix B of Ref. \cite{fgl}), including adopting the value for the systematic error level of $\epsilon=3\%$.

\begin{figure*}
    \includegraphics[width=3.3in]{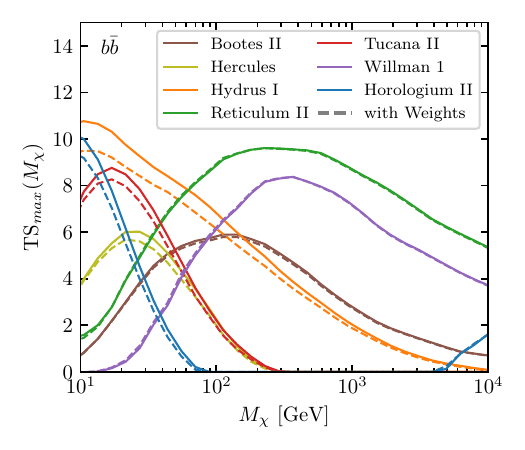}\hfill
    \includegraphics[width=3.3in]{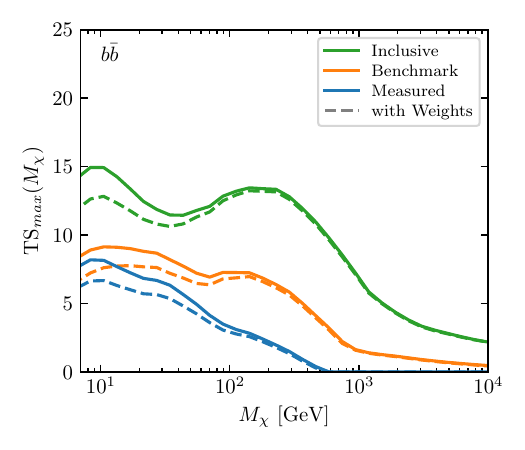}
    \caption{TS vs $M_{\chi}$ for the standard likelihood analysis (solid lines) and the weighted likelihood analysis (dashed lines) for the individual high TS dSphs ({\it left}), and the combined dSph samples ({\it right}).}\label{fig:wTSvM}
\end{figure*}

Figure \ref{fig:wTSvM} shows the comparison in the TS vs DM mass profiles between the standard likelihood analysis and the weighted likelihood analysis for individual dSphs as well as for the combined samples. Indeed, the weighted likelihood approach does mitigate to some extent the low-mass peaks, indicating that the relatively large TS values at low masses are at least somewhat attributable to background systematics. The same effect is also observed for the blank fields. Figure \ref{fig:TSvM_weights} shows the TS profile as a function of mass for the subsamples compared with the blank fields analyzed using the weighted likelihood method. Comparing with Figure \ref{fig:TSvM_sample} demonstrates the decrease in the blank field distribution at low masses as well as the dSph samples. Since the global significances are determined relative to the maximal TS values for all masses and channels, the decrease of the low-mass peak in the blank fields yields a slightly improved global significance for local excesses at masses where the effects of the weighted likelihood are not as prominent (e.g. the global significance for the Benchmark sample at the local significance peak improves from $0.5\,\sigma$ to $\sim 0.7\,\sigma$). The summary of local significance peaks for the weighted likelihood approach for both annihilation channels is presented in Table \ref{tab:weight_results}. We note that despite minor improvements to the dSphs significances, the implementation of the weighted likelihood analysis does not have any noticeable impact on the annihilation cross section upper limits in comparison to the standard likelihood approach.

\begin{figure*}
    \includegraphics[width=3.5in]{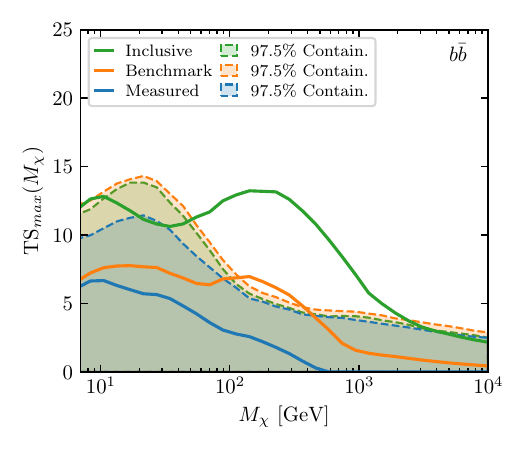}\hfill
    \includegraphics[width=3.5in]{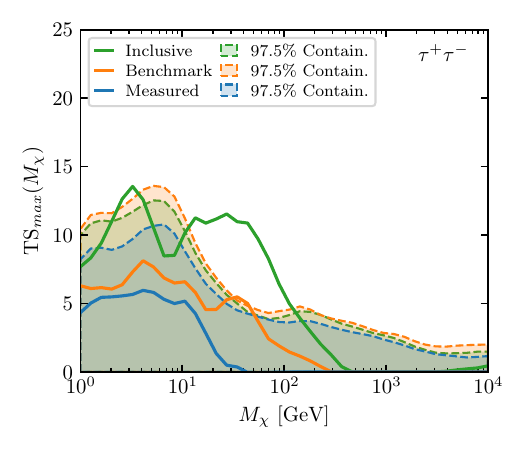}\\
        \caption{Maximum TS as a function of $M_{\chi}$ over all cross-section values for each dSph sample using the weighted likelihood method; shaded regions are the 97.5\% containment region for the combined blank fields. 
        \label{fig:TSvM_weights}
        }
\end{figure*}

\begin{table*}[tbph!]
\centering
\def\arraystretch{1.55}
\setlength{\tabcolsep}{4pt}

\begin{tabular}{lccc|ccc}

\hline\hline
 & \multicolumn{3}{c}{$b\bar{b}$} & \multicolumn{3}{c}{$\tau^+\tau^-$}\\
 \cline{2-4}  \cline{5-7}
 & $M_{\chi}$ &  $p_{\mathrm{local}}$ & $p_{\mathrm{global}}$  & $M_{\chi}$ &  $p_{\mathrm{local}}$ & $p_{\mathrm{global}}$
\\
 & [GeV]                    &                     &  & [GeV]                    &                     &
\\
\hline
Measured & 8.4 & $8.9\times 10^{-2}$ ($1.4\,\sigma$) & $2.0\times 10^{-1}$ ($0.8\,\sigma$) & 1.6 & $1.1\times 10^{-1}$ ($1.2\,\sigma$) & $2.7\times 10^{-1}$ ($0.6\,\sigma$)
\\
Benchmark & 180.5 &  $1.7\times 10^{-2}$ ($2.1\,\sigma$)&  $2.4\times 10^{-1}$ ($0.7\,\sigma$)  & 34.6 &  $2.2\times 10^{-2}$ ($2.0\,\sigma$) &  $3.1\times 10^{-1}$ ($0.5\,\sigma$)
\\
Inclusive & 289.4 & $6.0\times 10^{-4}$ ($3.2\,\sigma$) & $5.2\times 10^{-2}$ ($1.6\,\sigma$)  & 43.8 & $7.0\times 10^{-4}$ ($3.2\,\sigma$)  & $8.0\times 10^{-1}$ ($1.4\,\sigma$)
\\
\hline\hline
\end{tabular}
\caption{DM mass and local and global $p$ values (and significance) for each sample at the local significance peak using the weighted likelihood method.}\label{tab:weight_results}
\end{table*}

While this test demonstrates that the diffuse systematic can noticeably effect certain aspects of the analysis, the overall conclusions are minimally affected. For each subsample considered, changes in the local significance values and corresponding masses range from slight to negligible, as is the case for the derived upper limits on the annihilation cross-section.

\nocite{*}

\bibliography{ref}

\end{document}